\documentclass[useAMS,usenatbib]{mnras}
\usepackage{natbib}
\usepackage[utf8]{inputenc}
\usepackage{graphicx}
\usepackage{amsmath}

\newcommand{\Lsun}{\mbox{$L_{\odot}$}} 
 
\newcommand{\Rjup}{\mbox{$R_{\rm Jup}$}} 
\newcommand{\Mjup}{\mbox{$M_{\rm Jup}$}}
\newcommand{\meth}{\mbox{${\rm CH_4}$}}
\newcommand{\water}{\mbox{${\rm H_{2}O}$}} 
\newcommand{\teff}{\mbox{$T_{\rm eff}$}}
\newcommand{\lbol}{\mbox{$L_{\rm bol}$}}
\newcommand{\enstatite}{\mbox{${\rm MgSiO_3}$}}
\newcommand{\forsterite}{\mbox{${\rm Mg_2 SiO_4}$}}
\newcommand{\sodiumsulphide}{\mbox{${\rm Na_{2}S}$}}
\newcommand{\carbondioxide}{\mbox{${\rm CO_{2}}$}}

\title[Ross~458c]{The puzzle of the formation of T8 dwarf Ross~458c}
\author[Gaarn et al]{Josefine Gaarn$^{1}$\thanks{E-mail:
    j.jensen@herts.ac.uk}, Ben Burningham$^{1}$, Jacqueline K. Faherty$^{2}$, Channon Visscher$^{3,4}$,  Mark S. Marley$^{5}$, \newauthor Eileen C. Gonzales$^{2,6}$\thanks{51 Pegasi b Fellow}, Emily Calamari$^{7}$, Daniella Bardalez Gagliuffi$^{2}$, Roxana Lupu$^{8,9}$, 
    \newauthor Richard Freedman$^{8,10}$  \\ 
$^{1}$ Centre for Astrophysics Research, Department of Physics, Astronomy and Mathematics, University of Hertfordshire, Hatfield AL10 9AB \\
$^{2}$ Department of Astrophysics, American Museum of Natural History, New York, NY 10024, USA\\
$^{3}$ Department of Chemistry, Dordt University, Sioux Center, IA 51250, USA \\
$^{4}$ Space Science Institute, Boulder, CO 80301, USA \\
$^{5}$ Department of Planetary Sciences and Lunar and Planetary Laboratory, University of Arizona, Tucson, AZ, USA \\
$^{6}$ Department of Astronomy and Carl Sagan Institute, Cornell University, 122 Sciences Drive, Ithaca, NY 14853, USA \\ 
$^{7}$ The Graduate Center, City University of New York, New York, NY 10016, USA \\
$^{8}$ NASA Ames Research Center, Mail Stop 245-3, Moffett Field, CA 94035, USA \\ 
$^{9}$ Bay Area Environmental Research Institute, 625 2nd Street, Suite 209, Petaluma, CA 94952, USA\\
$^{10}$ SETI Institute, Mountain View, CA 94043, USA\\
}

\begin{document}
%
%  These Macros are taken from the AAS TeX macro package version 4.0.
%  Include this file in your LaTeX source only if you are not using
%  the AAS TeX macro package and need to resolve the macro definitions
%  in the BibTeX entries returned by the ADS abstract service.
%
%  If you plan not to use this file to resolve the journal macros
%  rather than the whole AAS TeX macro package, you should save the
%  file as ``aas_macros.sty'' and then include it in your paper by
%  using a construct such as:
%	\documentstyle[11pt,aas_macros]{article}
%
%  For more information on the AASTeX macro package, please see the URL
%	http://www.aas.org/publications/aastex.html
%  For more information about ADS abstract server, please see the URL
%	http://adswww.harvard.edu/ads_abstracts.html
%

% Abbreviations for journals.  The object here is to provide authors
% with convenient shorthands for the most "popular" (often-cited)
% journals; the author can use these markup tags without being concerned
% about the exact form of the journal abbreviation, or its formatting.
% It is up to the keeper of the macros to make sure the macros expand
% to the proper text.  If macro package writers agree to all use the
% same TeX command name, authors only have to remember one thing, and
% the style file will take care of editorial preferences.  This also
% applies when a single journal decides to revamp its abbreviating
% scheme, as happened with the ApJ (Abt 1991).

\def\aj{\rm{AJ}}                   % Astronomical Journal
\def\araa{\rm{ARA\&A}}             % Annual Review of Astron and Astrophys
\def\apj{\rm{ApJ}}                 % Astrophysical Journal
\def\apjl{\rm{ApJ}}                % Astrophysical Journal, Letters
\def\apjs{\rm{ApJS}}               % Astrophysical Journal, Supplement
\def\ao{\rm{Appl.~Opt.}}           % Applied Optics
\def\apss{\rm{Ap\&SS}}             % Astrophysics and Space Science
\def\aap{\rm{A\&A}}                % Astronomy and Astrophysics
\def\aapr{\rm{A\&A~Rev.}}          % Astronomy and Astrophysics Reviews
\def\aaps{\rm{A\&AS}}              % Astronomy and Astrophysics, Supplement
\def\azh{\rm{AZh}}                 % Astronomicheskii Zhurnal
\def\baas{\rm{BAAS}}               % Bulletin of the AAS
\def\jrasc{\rm{JRASC}}             % Journal of the RAS of Canada
\def\memras{\rm{MmRAS}}            % Memoirs of the RAS
\def\mnras{\rm{MNRAS}}             % Monthly Notices of the RAS
\def\pra{\rm{Phys.~Rev.~A}}        % Physical Review A: General Physics
\def\prb{\rm{Phys.~Rev.~B}}        % Physical Review B: Solid State
\def\prc{\rm{Phys.~Rev.~C}}        % Physical Review C
\def\prd{\rm{Phys.~Rev.~D}}        % Physical Review D
\def\pre{\rm{Phys.~Rev.~E}}        % Physical Review E
\def\prl{\rm{Phys.~Rev.~Lett.}}    % Physical Review Letters
\def\pasp{\rm{PASP}}               % Publications of the ASP
\def\pasj{\rm{PASJ}}               % Publications of the ASJ
\def\qjras{\rm{QJRAS}}             % Quarterly Journal of the RAS
\def\skytel{\rm{S\&T}}             % Sky and Telescope
\def\solphys{\rm{Sol.~Phys.}}      % Solar Physics
\def\sovast{\rm{Soviet~Ast.}}      % Soviet Astronomy
\def\ssr{\rm{Space~Sci.~Rev.}}     % Space Science Reviews
\def\zap{\rm{ZAp}}                 % Zeitschrift fuer Astrophysik
\def\nat{\rm{Nature}}              % Nature
\def\iaucirc{\rm{IAU~Circ.}}       % IAU Cirulars
\def\aplett{\rm{Astrophys.~Lett.}} % Astrophysics Letters
\def\apspr{\rm{Astrophys.~Space~Phys.~Res.}}
                % Astrophysics Space Physics Research
\def\bain{\rm{Bull.~Astron.~Inst.~Netherlands}} 
                % Bulletin Astronomical Institute of the Netherlands
\def\fcp{\rm{Fund.~Cosmic~Phys.}}  % Fundamental Cosmic Physics
\def\gca{\rm{Geochim.~Cosmochim.~Acta}}   % Geochimica Cosmochimica Acta
\def\grl{\rm{Geophys.~Res.~Lett.}} % Geophysics Research Letters
\def\jcp{\rm{J.~Chem.~Phys.}}      % Journal of Chemical Physics
\def\jgr{\rm{J.~Geophys.~Res.}}    % Journal of Geophysics Research
\def\jqsrt{\rm{J.~Quant.~Spec.~Radiat.~Transf.}}
                % Journal of Quantitiative Spectroscopy and Radiative Transfer
\def\memsai{\rm{Mem.~Soc.~Astron.~Italiana}}
                % Mem. Societa Astronomica Italiana
\def\nphysa{\rm{Nucl.~Phys.~A}}   % Nuclear Physics A
\def\physrep{\rm{Phys.~Rep.}}   % Physics Reports
\def\physscr{\rm{Phys.~Scr}}   % Physica Scripta
\def\planss{\rm{Planet.~Space~Sci.}}   % Planetary Space Science
\def\procspie{\rm{Proc.~SPIE}}   % Proceedings of the SPIE

\let\astap=\aap
\let\apjlett=\apjl
\let\apjsupp=\apjs
\let\applopt=\ao

\maketitle

\begin{abstract}
At the lowest masses, the distinction between brown dwarfs and giant exoplanets is often blurred and literature classifications rarely reflect the deuterium burning boundary. Atmospheric characterisation may reveal the extent to which planetary formation pathways contribute to the population of very-low mass brown dwarfs, by revealing if their abundance distributions differ from those of the local field population or, in the case of companions, their primary stars.    
The T8 dwarf Ross~458c is a possible planetary mass companion to a pair of M dwarfs, and previous work suggests that it is cloudy. We here present the results of the retrieval analysis of Ross~458c, using archival spectroscopic data in the 1.0 to 2.4 micron range. We test a cloud free model as well as a variety of cloudy models and find that the atmosphere of Ross~458c is best described by a cloudy model (strongly preferred). The \meth/\water\ is higher than expected at $1.97^{+0.13}_{-0.14}$. This value is challenging to understand in terms of equilibrium chemistry and plausible C/O ratios. Comparisons to thermochemical grid models suggest a C/O of $\approx 1.35$, if \meth\ and \water\ are quenched at  2000 K, requiring vigorous mixing. We find a [C/H] ratio of +0.18, which matches the metallicity of the primary system, suggesting that oxygen is missing from the atmosphere. 
Even with extreme mixing, the implied C/O is well beyond the typical stellar regime, suggesting a either non-stellar formation pathway, or the sequestration of substantial quantities of oxygen via hitherto unmodeled chemistry or condensation processes. 
 
\end{abstract}


%%%%%%%%%%%%%%%%%%%%%%%Introduction%%%%%%%%%%%%%%%%%%%%%%
\section{Introduction}
\label{sec:intro}

Brown dwarfs span the mass range between stars and giant exoplanets and are found as isolated objects, binary systems, and as companions to stars. Most discussions of the formation of brown dwarfs  consider them as the substellar extension of the star formation process \citep[e.g.][and references therein]{whitworth2007,StametellosWhitworth2009,Whitworth2018}. 
However, the growing sample of isolated planetary mass brown dwarfs \citep[e.g.][]{faherty2013,liu2013,schneider2016,theissen2018,bouy2022}, wide-orbit (resolved) planetary mass companions \citep[e.g.][]{faherty2021,milespaez2017,Burgasser2010}, and giant exoplanets with masses above the deuterium burning limit \citep[e.g.][]{bakos2009,rosenthal2021} raises the question as to the contribution of planetary formation pathways to the low-mass end of the substellar population.

The compositions of brown dwarfs may provide clues to their formation mechanism. For objects that are companions to stars, we might expect a shared carbon-to-oxygen (C/O) ratio if both formed at the same time and from the same material. Whereas a C/O ratio that differs might suggest an alternative formation path to that of the star. For example, very low-mass brown dwarfs may have formed like gas giant planets in disks, through either core accretion or gravitational instabilities \citep{BateBonnellBromm2002}. 
Different oxygen-bearing species condense at different locations, removing oxygen from the vapor and yielding a higher C/O ratio in the gas phase.  As a result, at the water ice line the C/O ratio of grains would decrease, and then again increase at the CO ice line \citep{Oberg2011}. An object formed in such a differentiated protoplanetary disk may thus inherit a C/O ratio that is quite unlike its parent star.  
\\
For the brown dwarf population that represents the substellar extension of the star formation process, we can expect the distribution of C/O ratios to follow that of the stellar field population. This distribution is generally tight and close to the solar value of 0.54 \citep{Asplund2005}, with a C/O $>$ 0.8 very unusual \citep[e.g.][]{Fortney2012,Nissen2013,Nakajima2016}. If a significant number of the the lowest-mass brown dwarf companions, on the other hand, formed via a planet-like formation scenario, we might expect their C/O distribution to be altered by comparison. 
\\

In this paper we present the analysis of the first target of a wider study of the local brown dwarf and substellar companion populations. Ross~458c was discovered by \citet{Goldman2010} and followed up by \citet{Burgasser2010} and \citet{Burningham2011}, and is a young T8 dwarf. Comparisons to evolutionary models suggest that it may be a planetary mass object, while its wide orbit of 1200~AU around an M dwarf binary argues against formation via core-accretion or disk fragmentation (if the object formed in situ).

Another unusual feature of Ross~458c is its apparent cloudiness \citep{Morley2012, Burgasser2010, Burningham2011}.
Clouds are a possible opacity source in brown dwarfs, as inferred from data, and theoretically expected from condensation of species possible at pressures and temperatures found in substellar objects \citep[see e.g. ][]{Kirkpatrick2005,Kirkpatrick2021,LoddersFegley2006}. Although T dwarfs have been generally thought to be cloud-free \citep[e.g.][]{Kirkpatrick2005}, it has been suggested that sulphide and alkali-salt clouds may start to appear in late-T dwarfs \citep{Morley2012}. 

In this paper we use retrieval analysis to investigate the composition and cloudiness of this interesting target. The retrieval technique follows on from and complements analysis using one dimensional, self-consistent, radiative-convective equilibrium ``grid models''. Grid models attempt to simulate the substellar atmosphere self-consistently using a handful of bulk parameters that typically include $T_{\rm eff}$,  $\log g$, metallicity, cloud parameters, and (sometimes) C/O \citep[e.g.][]{SaumonMarley2008,MarleyRobinson2015,MarleyNEW2021}. 
Also, vertical mixing can drive certain molecular abundances out of chemical equilibrium \citep[e.g.][]{Noll1997,Saumon2006,MarleyRobinson2015}. The extent to which this occurs depends on the mixing timescale, and this is often included via some additional parameter \citep[e.g. eddy diffusion coefficient $K_{zz}$, ][]{GriffithYelle1999, Saumon2006}.

A variety of techniques are used to fit these models to data to draw conclusions, from simple $\chi^2$ minimisation to more sophisticated Bayesian methods \citep[e.g.][]{Zhang2020}.
This can result in poor fits to the data, with the grid models inadequate to describe the data \citep{Line2017}. Due to their complexity, it is difficult to identify the poor assumption or missing physics that is driving the bad fit.

%The grid models can have inconsistencies between the different models and the data \citep{Patience2012}, and often assume solar abundances and thermochemical equilibrium, which is not always the case in the local field. 
%Another shortcoming of the grid models is the failure to consider the variability of the brown dwarf spectra \citep{Line2014}. 

Retrievals make fewer assumptions about the state of the atmosphere, and instead parameterise it over many more dimensions with little if any requirement for self-consistency. Whilst this can result in unphysical outcomes, which should be considered when interpreting the result, it also allows data-driven insights that can highlight, or work around,  shortcomings in self-consistent approaches \citep[e.g.][]{Burningham2021}. Another benefit of retrievals is that one gets a measure of the C/O ratio from the observations (here using \meth\ and~\water) as a retrieved parameter, as opposed to a prescribed ratio that is hardwired into the model.  
In this work we will use the {\it Brewster} retrieval code that has been demonstrated effectively in cloudy atmospheres in the substellar regime \citep{burningham2017,Gonzales2020,Burningham2021}.

%\citet{Line2015} built upon successful retrieval models previously applied to Earth's atmosphere, the atmospheres of solar system bodies, and to exoplanets (see \citet{Line2017} for examples). They demonstrated their retrieval framework on two brown dwarfs, Gliese 570D and HD3651b, whose primaries' abundances were known. Their retrieval framework was validated by showing that the abundances of the secondary dwarfs matched those of their primaries, assuming co-evolution \citep{Line2015}. 

The paper is organised as follows. In Section 2 we provide an overview of current literature on Ross~458c. Section 3 describes Brewster, the retrieval framework. In Section 4 we present the results of the retrievals. Section 5 is a discussion and analysis of the results, and Section 6 is the conclusion. 

%%%%%%%%%%%%%%%%%%%%%%%%%%Ross~458c%%%%%%%%%%%%%%%%%%%%%%%

\section{Ross~458 system}

\subsection{Ross~458AB}

The primary system consists of a M0.5 and M7, separated by a distance of 0.5" (about 5 AU) \citep{Beuzit2004}. The metallicity of the primary system is supersolar at [Fe/H] = +0.2 $\pm$ 0.05 \citep{Burgasser2010}. 

\citet{Burgasser2010} places constraints of the age of the Ross~458 system of 150 to 800 Myr, the upper limit based on the magnetic activity and H$\alpha$ emission of Ross~458A and the lower limit on the absence of low surface gravity indicators (such as the 7000 \AA \ KI doublet), which would be expected in young M dwarfs. 

\citet{Eggen1960} selected the system as a possible member of the Hyades open cluster, with an age of around 625 Myr \citep{Lebreton1997} whilst \citet{Nakajima2010} selected it as a possible member of the IC 2391 Moving Group, with an age of around 50 Myr. Any membership association would place further constraint of the age of the system. 

Table~\ref{tab:systemlitvals} summarizes previously published parameters for Ross~458A and Ross~458B.

%%%%%%The table of system parameter values os Ross458AB%%%%%%%%%%%%% 
\begin{center} 
\begin{table*} 
\begin{tabular}{c c c c}
\hline\hline
     Parameter & Value & Notes/units & Reference  \\
     \hline 
     $\alpha$ & 13:00:46.5802 & J2000 & \citet{Gaia}\\
     $\delta$ & +12:22:32.604 & J2000 & \citet{Gaia}\\
     $\pi$ & $86.8570 \pm 0.15$ & mas & \citet{Gaia} \\
     $\mu_{\alpha}$ & $-632.151 \pm 0.50$ & mas/yr & \citet{Gaia}\\
     $\mu_{\delta}$ & $-36.019 \pm 0.19$ & mas/yr & \citet{Gaia}\\
     Ross~458A Spectral type & M0.5 &  & \citet{Hawley1997} \\
     Ross~458B Spectral type & M7.0 &  & \citet{Beuzit2004} \\
     Period & 14.5 & yr & \citet{Heintz1994} \\
     Distance & $11.51 \pm 0.02$ & parsec & \citet{Gaia}\\
     Projected binary separation (minimum)$^{a}$ & 5.4 & AU & \citet{Beuzit2004}\\
     Binary Separation$^{a}$ & 0.475" & arcseconds & \citet{Beuzit2004}\\
     Projected separation (minimum) to Ross~458c$^{a}$ & 1168 & AU & \citet{Goldman2010}  \\
    Separation to Ross~458c$^{a}$ & 102" & arcseconds & \citet{Goldman2010}  \\
     \hline 
\end{tabular}
\caption{System literature values for the Ross~458 system. a) No uncertainties provided in the sources.
\label{tab:systemlitvals} }
\end{table*}
\end{center}

\subsection{Ross~458c}
Ross~458c is a T dwarf of spectral type T8 \citep{Burgasser2003} or T8.5p \citep{Burningham2011} and is a wide orbit companion to the binary Ross~458AB \citep{Goldman2010}. It is separated from its primaries by 102" (1200 AU). The system is at a distance of $11.5 \pm 0.2$~parsec \citep{Gaia}. 

\citet{Burgasser2010} found \teff~$=~650\pm 25$ K from fitting self-consistent grid models. This is in good agreement with the values found by \citet{Burningham2011} (using bolometric luminosity) of $695 \pm 60$~K. \citet{Filippazzo2015} found a similar value of $721 \pm 94$~K using their bolometric luminosity method. 

One of the key results in those previous examinations of Ross~458c was the somewhat surprising feature of clouds in a late-T dwarf. \citet{Burgasser2010} found that models which included cloud opacity were a better fit to the near-infrared spectrum of Ross~458c. They argued that it may indicate a resurgence of iron and silicate clouds that are thought to have disappeared at the low temperatures of the late-T dwarfs. 

\citet{Morley2012} investigated possible causes of the apparent cloud opacity. Their best fit model for Ross~458c was a 700 K, log g 4.0 model which incorporated sulphide (\sodiumsulphide, MnS and ZnS) clouds. \citet{Manjavacas2019} find spectrophotometric variability in Ross~458c with an amplitude at the $\approx 2$\% level, indicating a partial cloud cover rotating in and out of view. 
%Variability is generally associated with the L/T transition but is also found in T and Y dwarfs \citep{Morley2014} .  

%%%%%%%%%The table of parameter literature values of Ross458C%%%%%%%%%%%%%%%% 
\begin{center} 
\begin{table*}
\begin{tabular}{c c c c}
\hline\hline
     Parameter & Value & Notes/units & Reference  \\
     \hline 
     $\alpha$ & 13:00:41.15 & J2000 & \citet{Gaia}\\
     $\delta$ & 12:21:14.22 & J2000 & \citet{Gaia}\\
     Spectral type & T$8.9 \pm 0.3$ &  & \citet{Goldman2010} \\
     & T8 &  & \citet{Burgasser2010} \\
     & T8.5p &  & \citet{Burningham2011} \\
     $T_{\rm eff}$ & $695 \pm 60$ & K & \citet{Burgasser2010}\\
     & $650 \pm 25 $ & K & \citet{Burningham2011} \\
     & $804^{+30}_{-29}$ & K & \citet{Zhang2020}  \\
     & $721 \pm 94$ & K & \citet{Filippazzo2015} \\
     & $722^{+11}_{-12}$ & K & this work \\
     log g & 4-4.7 & ${\rm log(cm/s^2)}$ & \citet{Burningham2011} \\
     & 4 & ${\rm log(cm/s^2)}$ & \citet{Burgasser2010} \\
     & 3.7-4 & ${\rm log(cm/s^2)}$ & \citet{Morley2012} \\
     & $4.50 \pm 0.07 $ &  & this work \\
     $log_{10} L_{\rm bol}/\Lsun$ & $-5.62 \pm 0.03$ &  & \citet{Burgasser2010} \\
     & -5.61 &  & \citet{Burningham2011} \\
     & $-5.27 \pm 0.03 $ &  & this work \\
     Metallicity & $-0.06 \pm 0.20$ &  & \citet{Goldman2010} \\
      & 0 & assumed & \citet{Burgasser2010} \\
      & +0.3 & cloud-free & \citet{Burgasser2010} \\
      & 0 & assumed & \citet{Burningham2011} \\
      & 0.18 $\pm 0.04$ & [C/H] & this work  \\
     Age & $<$ 1 & Gyr & \citet{Manjavacas2019}\\
     & 0.4-0.8 & Gyr & \citet{West2008} \\
     & 0.15-0.8 & Gyr & \citet{Burgasser2010} \\
     Mass & 5-14 & \Mjup, inferred & \citet{Goldman2010} \\
     & 6.29-11.52 & \Mjup & \citet{Burgasser2010} \\
     & 5-20 & \Mjup, inferred & \citet{Burningham2011} \\
     & $2.3^{+2.3}_{-1.2}$  & \Mjup & \citet{Zhang2020} \\
     & $27^{+4}_{-4}$ & \Mjup & this work \\
     Radius & $1.19-1.29 $ & \Rjup & \citet{Burgasser2010}\\
     & $1.01-1.23$ & \Rjup & \citet{Burningham2011} \\
     & $0.68 \pm 0.06$  & \Rjup & \citet{Zhang2020} \\
     & $1.45 \pm 0.06$  & \Rjup & this work \\
     \hline 
\end{tabular}
\caption{Parameters literature values of Ross458c. 
\label{tab:parameterslitvals} }
\end{table*}
\end{center}

In more recent work, \citet{Zhang2020} attempted to estimate the parameters using the Sonora Bobcat grid models \citep{MarleyNEW2021}, using the prism spectra of Ross~458c. However they struggled to get a good fit. The model spectrum was brighter in the Y and J band, and had fainter flux in the blue wing of the H and K band. They argue that this is likely due to clouds, reduction in vertical temperature gradient, or CO/\meth\ or ${\rm N_2}$/${\rm NH_3} $ disequilibrium chemistry. As a result, the radius and inferred mass are small. The model atmospheres used in their study were cloud-free, and with a C/O fixed at solar. These limitations might result in inaccurate estimated parameters. This work will improve upon these shortcomings by having a larger set of less restrictive parameters.  

For the retrieval of Ross~458c, we make use of archival data, for full details of the data see \citet{Burgasser2010}. 
We have flux-calibrated the data using the same UKIDSS $J$~band photometry used in \citet{Burningham2011} to provide consistency between the two studies and remove a source of possible discrepancies between our results.

Table \ref{tab:parameterslitvals} provides an overview of the literature values of Ross~458c.  

%%%%%%%%%%%%%%%%%%%%Retrieval framework%%%%%%%%%%%%%%%%%%%

\section{Retrieval framework}
We use the Brewster retrieval framework, for which a more complete description can be found in \citep{burningham2017, Gonzales2020, Burningham2021}. The retrieval framework consists of the forward model, which produces the spectrum based on a set of parameters, and the retrieval model, which uses Bayesian inference to explore the posterior, estimate parameters and perform model selection. 
The following section contains a brief recap of key features, and updates for this work.  

\subsection{Forward model overview}
Brewster divides the atmosphere into 64 pressure layers with properties of temperature, gas opacity and cloud opacity. The atmosphere is also parameterised by gravity which sets the scale height. The output of the forward model is the flux at the top of the atmosphere, which must be scaled by the ratio of the squared radius, R, and distance to the target, D ($R^2 / D^2$).
As will be discussed in Section~\ref{sec:posteriorsampling}, we use two different algorithms for exploring the posterior: EMCEE \citep{ForemanMackie2013}, and nested sampling using PyMultiNest \citep{Buchner2014}. 
These two methods use slightly different parameterisations due to the differences in how the priors are treated in each algorithm. 
Specifically mass and radius are retrieved directly in the nested sampling version, whereas they are derived from the ($R^2 / D^2$) scaling parameter and $\log g$ in the EMCEE version. The thermal profiles also differ as will be discussion Section~\ref{sec:thermal}. \\

The forward model parameters are discussed below, and are listed in Table \ref{tab:priors} along with their priors. 

\begin{table*}
\begin{center}
\begin{tabular}{c c}
\hline
Parameter & Prior \\
\hline
gas fraction ($X_{gas}$) & log-uniform, $\log X_{gas} \geq -12.0$, $\sum_{gas}{X_{gas}} \leq 1.0$ \\
thermal profile$^{1}$:  $\alpha_{1}, \alpha{2}, P1, P3, T3$  & uniform on resulting $T$, $0.0~{\rm K} < T_{i} < 5000.0~{\rm K}$\\
thermal profile$^{2}$: $T_{\rm bot},T_{\rm botQ}, T_{\rm mid}, T_{\rm topQ}, T_{\rm top}$  &  uniform, $0.0~{\rm K} < T_{i} < 5000.0~{\rm K}$; $T_{\rm bot} < T_{\rm botQ} < T_{\rm mid} < T_{\rm topQ} < T_{\rm top}$\\ 
Mass$^{2}$ & uniform, $1 M_{\rm Jup} \leq R \leq 80 M_{\rm Jup}$\\
Radius$^{2}$ & uniform, $0.5 R_{\rm Jup} \leq R \leq 2.0 R_{\rm Jup}$\\ 
scale factor$^1$, $R^{2} / D^{2}$ & uniform, constrained by $0.5 R_{\rm Jup} \leq R \leq 2.0 R_{\rm Jup}$ \\
gravity$^{1}$, $\log g$ & uniform, constrained by $1 M_{\rm Jup} \leq gR^{2} / G \leq 80M_{\rm Jup}$\\
Power law cloud opacity ($\tau \propto \lambda^\alpha$), $\alpha$ & uniform, $-10 \leq \alpha \leq 10$ \\
cloud decay scale, $(\Delta \log_{10} P)_{decay}$  & uniform, $0 < (\Delta \log_{10}  P)_{decay} < \log_{10} P_{top} + 4$ \\
cloud thickness $(\Delta \log_{10} P)_{thick}$ & uniform, constrained by $\log_{10} P_{top} \leq \log_{10} P_{top} + (\Delta \log_{10} P)_{thick} \leq 2.3$\\
cloud total optical depth (extinction) at 1 $\mu$m$^{3}$ & uniform, $0.0 \leq \tau_{0} \leq 100.0$ \\
Hansen distribution effective radius, $a$ & log-uniform, $-3.0 < \log_{10} a {\rm (\mu m)} < 3.0$ \\
Hansen distribution spread, $b$ & uniform, $0 < b < 1.0$ \\
Wavelength shift & uniform, $-0.01< \Delta \lambda < 0.01 \micron$  \\
tolerance factor, $o$ & uniform, $\log (0.01 \times min(\sigma_{i}^{2})) \leq o \leq \log(100 \times max(\sigma_{i}^{2}))$ \\

\hline
\end{tabular}
\caption{Priors for retrieval model. Notes: 1) EMCEE only; 2) nested sampling only. The tolerance factor, $o$ is to allow to unaccounted sources of uncertainty. The wavelength shift is the shift in wavelength between the data and the model. See \citet{burningham2017} for a complete description. The cloud decay scale, $(\Delta \log_{10} P)_{decay}$, is constrained within the height of the atmosphere by $\log_{10} P_{top} + 4$. The cloud thickness, $(\Delta \log_{10} P)_{thick}$, is constrained by 2.3, the bottom of the atmosphere.
\label{tab:priors}
}
\end{center}
\end{table*}

\subsubsection{Thermal structure} 
\label{sec:thermal}
As is typical in atmospheric retrievals, we do not directly retrieve the temperature for all 64 layers in our model atmosphere, but instead use a smaller number of parameters to set the thermal profile.  
In previous works \citep[e.g.][]{burningham2017, Burningham2021,Gonzales2020}, we have used the \citet{MadhusudhanSeager2009} parameterisation for the thermal profile.This scheme treats the atmosphere as three zones: 

\begin{equation}
\begin{aligned}
P_{0} < P < P_{1}: P  = P_{0} e^{\alpha_{1}(T - T_{0})^{\frac{1}{2}}}   \hfill (\text{Zone 1})\\
P_{1} < P < P_{3}: P  = P_{2} e^{\alpha_{2}(T - T_{2})^{\frac{1}{2}}}   \hfill (\text{Zone 2})\\
P > P_{3} : T = T_{3}  \hfill (\text{Zone 3})
\end{aligned}
\label{eqn:madhu}
\end{equation}
where $P_{0}$, $T_{0}$ are the pressure and temperature at the top of the atmosphere, which becomes isothermal with temperature $T_{3}$ at pressure $P_{3}$.
In its most general form, a thermal inversion occurs when $P_{2} > P_{1}$. Since $P_{0}$ is fixed by our atmospheric model, and continuity at the zonal boundaries allows us to fix two parameters, we consider six free parameters $\alpha_{1}$, $\alpha_{2}$, $P_1$, $P_2$, $P_3$, and $T_3$.  If we rule out a thermal inversion by setting $P_{2} = P_{1}$ \citep[see Figure 1, ][]{MadhusudhanSeager2009},  we can further simplify this to five parameters $\alpha_{1}$, $\alpha_{2}$, $P_1$, $P_3$, $T_3$.

For our new nested sampling version of Brewster we have employed a simple 5 point T-P profile, which sets temperatures at the base ($T_{bot}$), top ($T_{top}$), mid-point ($T_{mid}$ and quarter pressure depths ($T_{topQ}$, $T_{botQ}$) in the atmosphere in $\log_{10} P$. These points are then joined by spline-interpolation to define the 64 layer temperatures. The only restrictions on the values these 5 points may take is that they lie in the range $0 < T < 4000$~K, and decrease towards shallower pressures.

\subsubsection{Gas Opacity}
The gas opacity is set by the gas fractions in each layer. In this work we directly retrieve these as vertically constant mixing ratios. Previous retrievals of late-T~dwarfs using the assumption of chemical equilibrium to allow for vertically varying abundances found such models poorly ranked \citep[see e.g. ][]{Gonzales2020}. 
Moreover, grid model fits in \citet{Burningham2011} suggested that non-equilibrium chemistry is important for this Ross~458c. 
As such, we have not performed any retrievals under the assumption of chemical equilibrium in this work. 
%or with vertically varying abundances calculated on the assumption of thermochemical equilibrium. The latter method retrieves [Fe/H] and C/O, as opposed to individual gas abundances. The gas fractions in each layer are interpolated from a table of equilibrium abundances pre-computed as a function of T, P, [Fe/H] and C/O \citep{Visscher2006,Visscher2010,Visscher2012,MarleyNEW2021}. 

\subsubsection{Opacity data}
In this work we consider the absorbing gases \water, \meth, CO, \carbondioxide, ${\rm NH_{3}}$, ${\rm H_{2}S}$, Na, K.
Their opacities are from the compendium \citep{Freedman2008} and \citep{Freedman2014}. 
For the Na and K D1 and D2 lines, broadened by collisions with ${\rm H_2}$ and He, the Lorentzian line profile becomes inadequate, and line wing profiles are taken from \citep{Allard2007a, Allard2007b, Allard2016}. We also consider the alkali opacities from \citet{BurrowsVolobuyev2003} as an alternative given the uncertainty surrounding the best treatment for this issue, as there is no agreement in literature as to which is the preferred choice. \citet{Gonzales2020} tested both Burrows and Allard alkalis on their L7+T7.5p SDSS J1416+1348AB binary target. The 1416A component (L7) was best fit using the Allard treatment, whilst the 1416B 
component (T7.5p) was best fit using the Burrows treatment. The Allard alkalis provide consistent abundances between the co-evolving pair, and is hence the best fit for the pair as a whole. We follow \citet{Gonzales2020} in testing both the Burrows and the Allard treatment for alkalis. 

We include continuum opacities for H$_{2}$--H$_2$ and H$_2$--He collisionally induced absorption, using cross sections from \citet{Richard2012, Saumon2012}. We also include Rayleigh scattering due to H$_{2}$, He and CH$_{4}$, and continuum opacities due to bound-free and free-free absorption by H$^-$ \citep{John1988, Bell1987} and free-free absorption by H$^{-}_{2}$ \citep{Bell1980}.

\subsubsection{Cloud opacity}
\label{sec:cloudmodel}
The cloud opacity is parameterised in several ways, which have between 3 and 5 parameters. These parameters fall broadly into two categories: those that describe how the cloud opacity is distributed in the atmosphere and those that describe the cloud's optical properties. \\

Cloud structures: The clouds' distribution in the atmosphere is parameterised as either a ``deck" or ``slab" cloud. In the deck cloud, it is not possible to see the bottom of the cloud. The deck cloud has the following parameters: 
\\
1) Pressure at which the optical depth passes 1, looking down
\\
2) The decay height, the pressure over which the optical depth drops by a factor of 2 
\\

In contrast to the deck cloud, in the slab cloud, it is possible to see the bottom of the cloud. The slab cloud has the following parameters: 
\\
1) The total optical depth of the cloud at 1 \micron\  (reached at the base of the cloud)
\\
2) Pressure at the top of the cloud (where $\tau_{cloud} = 0$)
\\
3) The thickness of the cloud slab (in $d\log P$)
\\

 Cloud optical properties: The cloud opacity can be parameterised in one of three ways. The optical depth is either grey (wavelength independent), a power law ($\tau = \tau_0 \times \lambda^\alpha$) where $\tau_0$ is the optical depth at 1 \micron, or Mie scattering. If the cloud opacity is power law dependent as opposed to grey, an extra parameter is added, the power index $\alpha$, defining the wavelength dependent cloud opacity as $\tau \propto \lambda^\alpha$. In both the grey and the power-law case, scattering is neglected, and the cloud is assumed to be absorbing. When using Mie scattering, we have two additional parameters, the Hansen \textit{a} and \textit{b} parameters \citep{Hansen1971}. The Hansen \textit{a} parameter is the particle size distribution effective radius in log(r/\micron). Hansen \textit{b} parameter is the particle size distribution spread. Therefore, for a deck cloud, we then have 4 parameters, and for the slab 5. 
\\
A variety of condensates are potentially able to form in cool T dwarf atmospheres and contribute to the cloud opacity. \citet{Burgasser2010} hypothesised a reemergence of iron and silicate clouds for late-T dwarfs, which would have broken up at the L-T transition. 
Of the iron and silicate clouds, it is the silicate clouds that will be found at the lowest temperatures and shallowest pressures. 
These silicate clouds are expected to be composed of enstatite (\enstatite), forsterite (\forsterite) or ${\rm SiO}_2$ \citep[quartz, e.g. ][]{LoddersFegley2006,Visscher2010,Burningham2021}. 
Since we are unlikely to be able to distinguish between these species using only near-infrared data we have only tested one: \enstatite. 
\citet{Morley2012} argues that sulphide clouds may be present in late-T dwarfs and tests \sodiumsulphide, MnS and ZnS, as they are expected to condense at lower temperatures. The abundance of ZnS is low and they find that it does not create an observable change in the spectrum; its optical depth is negligible, so ZnS is not included in the models we test here. KCl is also expected to condense at cooler temperatures \citep{Morley2012}, and is one of the models tested here. Graphite clouds can appear in Carbon-rich atmospheres with super-solar C/O ratio \citep{Moses2013, Moses2013B}. As a proxy for graphite, we used optical data for soot, due to its high carbon content.

Only one cloud type is tested per model run. We don't expect spectral features arising from Mie scattering of particular condensate compositions 
 from any of the expected cloud species to be  observable in the $1.0-2.5~\mu$m region.
So, we judged that the additional computational load and complexity due to adding extra model combinations in the form of multiple cloud models, would not be justified in terms of improvements to the fit.
\bigbreak
 We calculate scattering and absorption Mie coefficients from real and imaginary refractive indices for our condensate species obtained from the sources of optical data shown in Table \ref{tab:optics}. For grey and power-law opacity, no optical data are used.

\begin{table}
\begin{center}
\begin{tabular}{c c}
\hline
Condensate & Reference \\
\hline
MnS & \citet{HuffmanWild1967} \\
KCl & \citet{Querry1987} \\
\enstatite & \citet{ScottDuley1996} \\
\sodiumsulphide & \citet{Khachai2009} \\
soot & \citet{Dalzell1969} \\
\hline
\end{tabular}
\caption{Sources for optical data for condensates used in this work
\label{tab:optics}
}
\end{center}
\end{table}

\subsection{Posterior sampling} \label{sec:posteriorsampling}

In previous works, Brewster used the EMCEE sampler \citep{ForemanMackie2013}.  
However, this method of sampling has some drawbacks such as: uncertainty surrounding convergence; the challenge of exploring the entire probability volume to avoid sensitivity to initial values due to local maxima; and difficulty surrounding the calculation of the Bayesian Evidence for model selection. 
For this work, Brewster has been updated to also include nested sampling by incorporating the PyMultiNest sampler \citep{Buchner2014}, which avoids these issues. We test both the EMCEE and PyMultiNest posterior sampling techniques to check for consistency between the two methods.  

%Nested sampling samples the entire parameter space, and can prevent the walkers from being stuck in a local minimum or maximum. 

\subsection{Model selection}
The EMCEE sampler does not allow for the evaluation of the Bayesian evidence, but provides a maximum likelihood, which is used to calculate the Bayesian Information Criterion (BIC). 
 The model with the lowest BIC is the preferred model. The strength of the preference of one model over another is defined by \citet{KassRaferty1995} as such shown in Table~\ref{tab:deltabic}. 

\begin{table}
\begin{center}
\begin{tabular}{c c}
\hline
$\Delta \textrm{BIC}$ & Strength \\ 
\hline
0 to 2 & no preference worth mentioning \\
2 to 6 & positive \\
6 to 10 & strong \\
${>10}$ & very strong \\
\hline
\end{tabular}
\caption{Bayesian Information Criterion. 
\label{tab:deltabic}
}
\end{center}
\end{table}

%\begin{itemize}
%\item $0 < \Delta \textrm{BIC} <2$: no preference worth mentioning
%\item $2 < \Delta \textrm{BIC} <6$: positive
%\item $6 < \Delta \textrm{BIC} <10$: strong
%\item $\Delta \textrm{BIC} >10$: very strong
%\end{itemize}

The PyMultiNest sampler returns the nested importance sampling global log-evidence  \citep{Feroz2019}. 
%The nested importance sampling is a summation of the Bayesian evidence, and is preferred over nested sampling as it does not discard points that would have been discarded by the regular nested sampling due to constrained likelihood sampling. 
For two competing models, model selection between models 1 and 0 of the preferred model is achieved as follows: 

\begin{equation}
R=\frac{\mathcal{Z}_{1}}{\mathcal{Z}_{0}} \frac{Pr(H_{1})}{Pr(H_{0})}
\end{equation}
 
$\mathcal{Z}$ is the Bayesian evidence. Pr($H_{1}$)/Pr($H_{0}$) is the prior probability ratio for the two models, which can in most cases, including this, be set to unity. The preferred model is the model with the larger value for $\mathcal{Z}$, with the strength of the preference of one model over another being the ratio of their evidences \citep{Feroz2019}. The MultiNest algorithm returns the log-evidence (logEv), and hence the strength of the preference is defined as the difference between the models' logEv ($\Delta$logEv). 
Narrative statements corresponding to boundaries in the values for $\Delta$logEv are defined by \citet{KassRaferty1995} as shown in Table \ref{tab:bayes}. 

\begin{table}
\begin{center}
\begin{tabular}{c c}
\hline
$\Delta$logEv & Strength \\
\hline
0 to 0.5 & no preference worth mentioning \\
0.5 to 1 & substantial \\
1 to 2 & strong \\
$> 2 $ & decisive \\
\hline
\end{tabular}
\caption{Bayesian evidence difference. 
\label{tab:bayes}
}
\end{center}
\end{table}

%%%%%%%%%%%%%%%%%%%%%%%%Results%%%%%%%%%%%%%%%%%%%%%%%%%
\section{Results}

\subsection{Summary}
We tested both the EMCEE and the PyMultiNest options for sampling the posterior, to confirm consistency between results. The results quoted here are from PyMultiNest, as it is preferred due to sampling the entire parameter space, avoiding being stuck in local minima and maxima. The best fit model for the nested sampling is a cloudy model, with a power-law slab cloud. For nested sampling we use the Bayes factor to compare the different models. The evidence difference to the winning model is listed in Table~\ref{tab:multinestresults}. The evidence difference to the second ranked model is substantial. 

For the EMCEE sampling, the top ranked model is the power-law slab cloud, with the alternative treatment of the alkali line broadening. The second ranked model is the power-law slab cloud. Both PyMultiNest and EMCEE have the same two models ranked in top, the order is just switched. For EMCEE we use the $\Delta$BIC for model selection. The $\Delta$BIC is shown in Table~\ref{tab:emceeresults}. There is a strong preference for the top ranked model as opposed to the second ranked, and a very strong preference against the lower ranked models. 

The top-ranked models are cloudy models with the cloud opacity as a power-law. ``Real" clouds with Mie-scattering are rejected, as is the cloud-free model. 

We consider two different alkali profiles. For the PyMultiNest sampler, the Allard alkalis are ranked second, whereas for the EMCEE sampler, the Allard alkali profiles are ranked first, with the Burrows profile second. We cannot assert which is the right profile to use, as it is dependent on methodology.

\begin{center} 
\begin{table*}
\begin{tabular}{c c c c}
\hline\hline
     Model type & Nparams  & $\textrm \meth/\water$ & log-Evidence Difference  \\
     \hline 
     Power law slab cloud & 20 &  $1.97^{+0.13}_{-0.14}$ & 0\\
     Power law slab cloud, Allard alkalis & 20 & $2.13^{+0.14}_{-0.15}$ & -0.83\\
     \enstatite\ slab cloud & 21 & $1.86^{+0.19}_{-0.18}$ & -8.29 \\
     \sodiumsulphide\ slab cloud & 21 & $1.82^{+0.20}_{-0.19}$ & -8.80 \\
     Cloud free & 16 & $1.74^{+0.20}_{-0.19}$ & -9.27\\
     Grey deck cloud & 18 & $1.68^{+0.20}_{-0.20}$ & -10.85 \\
     KCl deck cloud & 20 & $1.79^{+0.17}_{-0.18}$ & -12.05\\
     \enstatite\ deck cloud & 20 & $1.76^{+0.19}_{-0.18}$ & -12.08 \\
     Power law deck cloud & 19 & $1.70^{+0.20}_{-0.21}$ & -12.31 \\
     Soot deck cloud & 20 & $1.32^{+0.28}_{-0.23}$ & -13.57 \\
     KCl slab cloud & 21 & $1.73^{+0.18}_{-0.17}$ & -13.88 \\
     Grey slab cloud & 19 & $1.63^{+0.17}_{-0.16}$ & -14.40 \\
     MnS slab cloud & 21 & $1.71^{+0.17}_{-0.17}$ & -14.49 \\
     MnS deck cloud & 20 & $1.75^{+0.19}_{-0.21}$ & -18.20 \\
     Soot slab cloud & 21 & $1.45^{+0.23}_{-0.37}$ & -18.37 \\
     %\sodiumsulphide deck cloud & 20 & 1.26 & -21.34\\
     \hline 
\end{tabular}
\caption{List of models tested in this work for Ross~458c, along with the log-evidence difference, using the PyMultiNest sampler. 
\label{tab:multinestresults} }
\end{table*}
\end{center}

\begin{center} 
\begin{table*} 
\begin{tabular}{c c c c}
\hline\hline
     Model type & Nparams  & $\textrm \meth/\water$ & $\Delta$BIC  \\
     \hline 
     Power law slab cloud, Allard alkalis & 21 & $1.82^{+0.31}_{-0.46}$ & 0\\
     Power law slab cloud & 21 & $1.69^{+0.31}_{-0.51}$ & +6\\
     Grey slab cloud & 20 & $1.67^{+0.28}_{-0.56}$ & +25 \\
     Soot slab cloud & 21 & $1.79^{+0.22}_{-0.24}$ & +27 \\
     \sodiumsulphide\ slab cloud & 21 & $1.83^{+0.19}_{-0.23}$ & +35 \\
     MnS slab cloud & 21 & $1.71^{+0.24}_{-0.29}$ & +44\\
     \enstatite\ slab cloud & 21 & $1.42^{+0.37}_{-0.42}$ & +51 \\
     KCl slab cloud & 21 & $1.28^{+0.32}_{-0.21}$ & +68\\
     Cloud free & 16 & $1.66^{+0.23}_{-0.25}$ & +81 \\
     Power law deck cloud & 20 & $1.61^{+0.36}_{-0.80}$ & +84 \\
     MnS deck cloud & 20 & $1.66^{+0.32}_{-0.64}$ & +87 \\
     \sodiumsulphide deck cloud & 20 & $1.76^{+0.18}_{-0.19}$ & +91\\
     Grey deck cloud & 19 & $1.57^{+0.32}_{-0.57}$ & +92 \\
     \enstatite\ deck cloud & 20 & $1.71^{+0.41}_{-1.06}$ & +92 \\
     KCl deck cloud & 20 & $1.71^{+0.27}_{-0.31}$ & +96 \\
     Soot deck cloud & 20 & $1.52^{+0.26}_{-0.23}$ & +100\\
     \hline 
\end{tabular}
\caption{List of models tested in this work for Ross~458c, along with the $\Delta$BIC, using the EMCEE sampler. 
\label{tab:emceeresults}}
\end{table*}
\end{center}

\subsection{Retrieved spectrum}
\begin{figure}
\hspace{-0.8cm}
    \includegraphics[width=270pt]{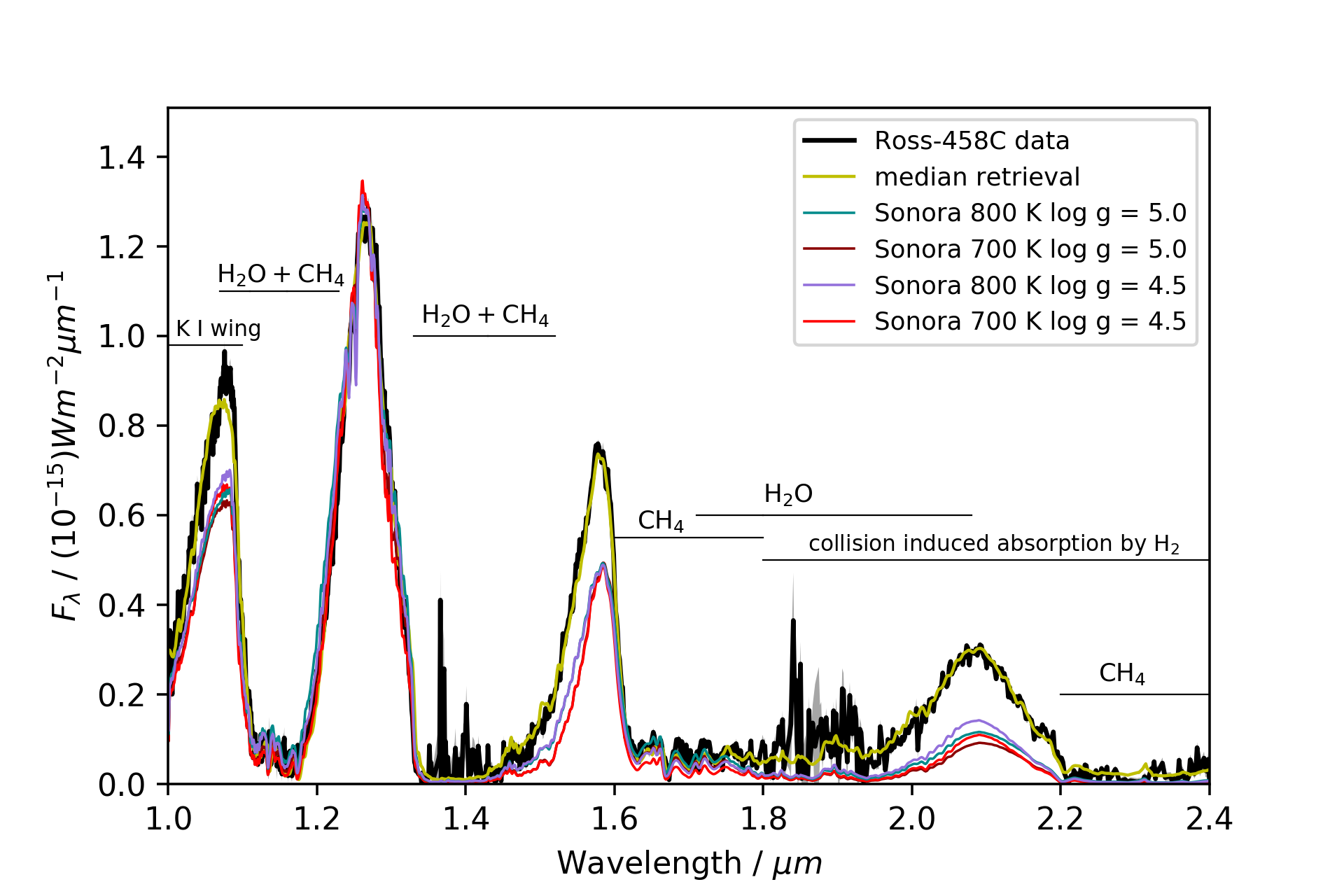}
    \caption{Spectrum of Ross~458c in the 1-2.4 $\mu$m range with retrieved spectrum of winning model (power-lab slab cloud) in green, and errors in grey. Over-plotted are self-consistent cloud-free Sonora grid models. The models are scaled to match the $J$~band flux.}
    \label{fig:spectrum}
\end{figure}

Figure~\ref{fig:spectrum} shows the spectrum with cloud-free Sonora models over-plotted, of \teff~$= 800$~K and $700$~K, and $\log g = 4.5$~and~5.0, and the main absorption features. The models are all ``normalised" to the $J$~band. All of the plotted grid models underestimate the flux in the $Y$, $H$, and $K$ bands compared to the data for Ross~458c. 
This may indicate that one of the key issues for the grid models is that their predicted flux in the $J$~band is too bright relative to the rest of the near-infrared spectrum. 

 Our retrieval model provides a much better fit to the data across the entire near-infrared wavelength range covered. This likely reflects the greater flexibility available to a retrieval model, and its ability to allow for non-solar abundance ratios. 
A key feature in our ability to fit the $J$~band peak is the inclusion of clouds (as will be discuss in Section~\ref{sec:cloudprops}).
The only region that the retrieval model struggles to fit well is the $Y$~band peak, and this likely reflects challenges associated with the pressure-broadened wings of the alkali D-lines, which is shared by the grid models.

\subsection{Thermal profile}
\begin{figure}
    \centering
    \includegraphics[width=280pt]{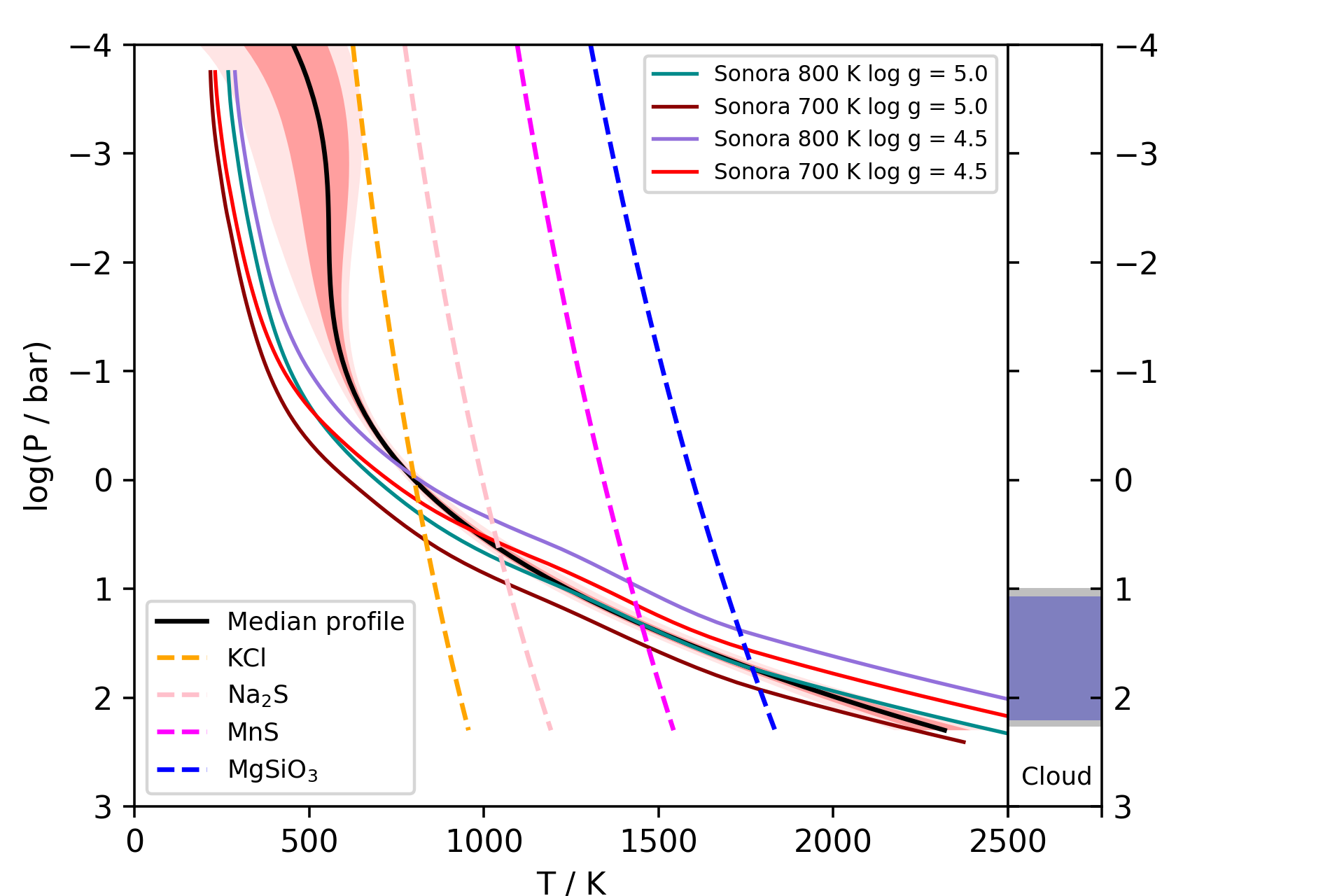}
    \caption{Pressure-temperature profile of winning model (power-lab slab cloud). The dashed lines are condensation curves for the different clouds tested. The curves do not mean that the particular cloud will condense, it just means that it can condense at that pressure and temperature.  The right part of the figure is the placement of the cloud in the atmosphere, the grey shaded parts are the 1$\sigma$ errors. The solid coloured lines are the self-consistent Sonora Bobcat grid models.
    \label{fig:profile} 
    }
\end{figure}

Figure~\ref{fig:profile} shows the thermal profile for the winning model along with comparison to self-consistent grid models and cloud condensation curves. The placement of the cloud in pressure space is plotted on the side. The thermal profiles for comparison are the Sonora Bobcat models \citep{MarleyNEW2021} with \teff = 700 K and 800 K and log g = 4.5 and 5.0. The comparison models were selected based on the retrieved \teff\ and log g. 

Generally, the models match the retrieved profile best at deeper pressures, where they follow the retrieved gradient well. The model that matches the retrieved profile best at deeper pressures is the \teff = 800 K, log g = 5.0. The models do not match closely at shallower pressures, where the differences are significant, as our retrieved profile is warmer (see Figure~\ref{fig:profile}).  
At the shallowest pressures in our model ($\log (P / {\rm bar}) \sim -3 - -4$) there is little contribution to the flux (see Figure~\ref{fig:contribution_function2}), and a correspondingly large scatter in the retrieved temperature of atmosphere. 
However, at slightly deeper pressures ($\log (P / {\rm bar}) \sim -1 - -2$) there is significant contribution of flux, and the retrieved thermal profile is quite tightly constrained and diverging from the grid models with a more isothermal gradient. 

This difference between the retrieved profile and other grid models at shallow pressures is also seen in other works, such as the warmer L dwarfs in \citet{Burningham2021}, and in works such as \citet{Line2017} sample of late-T dwarfs. It is not seen in \citet{Gonzales2020} L7+T7.5p pair, nor in Gliese 570D  \citep[]{Line2015,burningham2017}.

%Figure~\ref{fig:thermalcomparison} shows the comparison of the thermal profiles between the models. All the profiles of the tested models are similar, both at high and low pressures, indicating that the thermal profile is not too dependent on cloud model. 

%\begin{figure}
%    \centering
%    \includegraphics[width=280pt]{thermalcomparison.png}
%    \caption{Comparison of median thermal profiles of all models tested. The winning model in black, and other models in green.
%    \label{fig:thermalcomparison} 
%    }
%\end{figure}

\subsection{Cloud properties}
\label{sec:cloudprops}
In agreement with previous works \citep{Burgasser2010, Burningham2011, Morley2012}, we find that the atmosphere of Ross~458c is best described with a cloudy model. The winning model is a cloudy one, structured as a slab cloud. \citet{Burgasser2010}, using grid models, found that enstatite cloud models provided a good fit to the data, whereas \citet{Morley2012}, though not testing enstatite clouds found that sulphide clouds fit the data best. Model selection rejects ``real" clouds with Mie scattering, so we are unable to distinguish the cloud species based on optical properties.

Figure~\ref{fig:profile} shows the condensation curves for the tested cloud species alongside our retrieved thermal profile. The curves do not mean that the particular cloud {\it will} condense, but rather they indicate where it {\it can} condense. So, a given cloud species {\it can} condense to the left (cooler) of its condensation curve on Figure~\ref{fig:profile}. 
Two clouds that could condense at our retrieved cloud location are the MnS and \enstatite\ clouds. KCl and \sodiumsulphide\ condense at cooler temperatures. 
So, our thermal profile and the placement of the cloud deep in the atmosphere is suggestive of enstatite or other silicate clouds due to the retrieved cloud location with respect to the condensation curves. 
We note that the other silicates that were not tested (see Section~\ref{sec:cloudmodel}) have very similar condensation temperatures, so this cloud may actually be a mix of these very similar silicates. This will be difficult to test since these clouds appear to be so deep in the atmosphere. JWST data will likely incorporate the spectral features around $8 - 10\ \mu$m arising from Mie scattering of these silicates. However, at those wavelengths clouds are likely to be well below the photosphere.

\begin{figure}
\hspace{-0.8cm}
    \includegraphics[width=290pt]{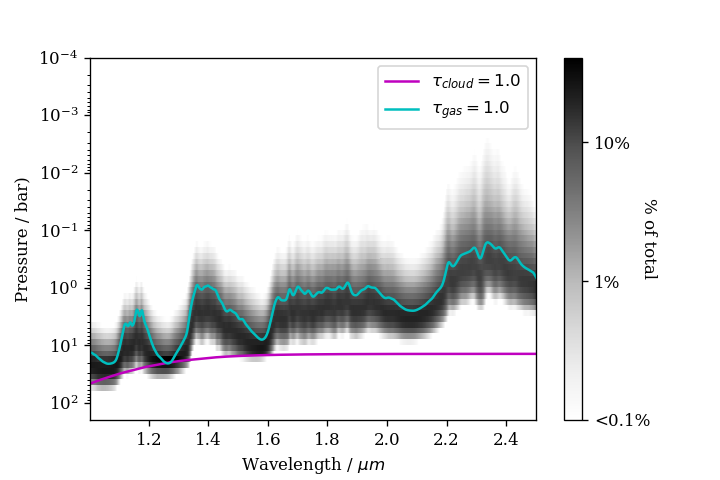}
    \caption{Contribution function of winning model (power-law slab cloud) with $\tau_{cloud} = 1$ in purple and $\tau_{gas} = 1$ in blue.}  
    \label{fig:contribution_function2}
\end{figure}

Figure~\ref{fig:contribution_function2} shows the contribution function. This shows where in the atmosphere the flux is coming from. The shading of the area indicates the relative contribution of flux from that layer. The darker shaded the area, the more flux from that layer. The blue line is the gas opacity where  $\tau_{gas}$ = 1.0. The cloud is the purple line (at $\tau_{cloud}$ = 1.0), and is placed in the bottom of the atmosphere, below the photosphere, except for an overlap at the $J$~band, at around 1.3 $\mu$m. 

This suggests that the cloud has the biggest impact in the $J$~band peak, where it reduces the flux from the deep atmosphere that escapes between the deep water and methane absorption bands on either side. 
This is what drives the much improved fit for our retrieval model compared to the cloud free grid models shown in Figure~\ref{fig:spectrum}. 
%As is usual, we have normalised this plot to the $J$~band peak. The grid models' poor apparent fit to the K band is actually driven by their overly bright J band, due to lack of clouds. 

\subsection{Gas abundances}
Figure~\ref{fig:postcorner} shows the retrieved bulk properties of Ross~458c, including the retrieved values for the gas abundances, along with $\log g$. Some gases, such as CO, \carbondioxide, and H$_{2}$S, are not well constrained, as seen on the histogram distribution. This has been the case for all near-infrared late-T dwarf retrievals using low resolution near-infrared spectra \citep{Line2017, Gonzales2020}.
To date, only \citet{Tannock2022} have detected H$_{2}$S in a near-infrared T~dwarf spectrum, and it required spectroscopy with $R\sim 45000$, 100x higher than used here. Even at this high-resolution, CO is still invisible due to masking by stronger CH$_{4}$ absorption.

We do detect and constrain the abundances of 
\water, \meth, ${\rm NH_3}$, and Na+K. This is the same set of molecules constrained by other works in this temperature regime \citep{Line2017, Zalesky2022}.

The corner plot also shows the C/O (here using the \meth/\water\ ratio), of 1.97, the calculation of which will be discussed in the Section~\ref{sec:analysis}. 

\begin{figure}
    \centering
    \includegraphics[width=225pt]{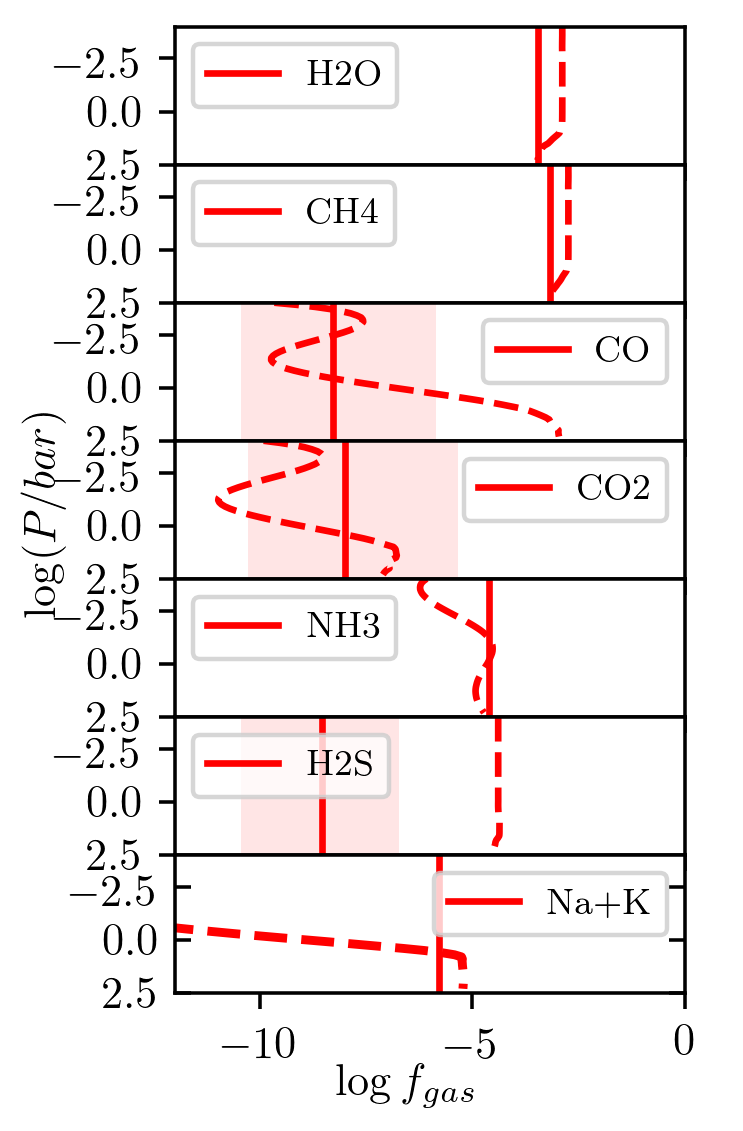}
    \caption{Median retrieved gas fractions compared to equilibrium predictions for $[M/H] = 0.2$ which matches Ross~458AB \citep{Burgasser2010}. The assumed C/O~$= 1.35$ is the highest available in the thermochemical grid. 
    The solid red lines are the retrieved gas fractions with the shaded are the 16th to 84th percentiles, and the dashed lined are the equilibrium predictions.
    \label{fig:NEWabundance}}
\end{figure}

The abundance plot shown in Figure~\ref{fig:NEWabundance} shows the retrieved gas mixing ratios along with equilibrium predictions from our thermochemical grid models \citep{Visscher2006,Visscher2010,Visscher2012,MarleyNEW2021}.
  
Our retrieved abundance for ${\rm NH_3}$ is consistent with chemical equilibrium predictions, whilst CO and \water\ are not. The abundance profiles for \water\ and \meth\ are close to vertical, somewhat justifying our vertically constant mixing ratio assumption in the retrieval model.  
\water\ and \meth\ match their abundances at 100 bars (2000 K),  and this may be suggestive of their chemistry being quenched at this level due to rapid vertical mixing.  This is discussed further in Section~\ref{sec:noneq}.

\begin{figure*}
\hspace{-0.8cm}
\includegraphics[width=550pt]{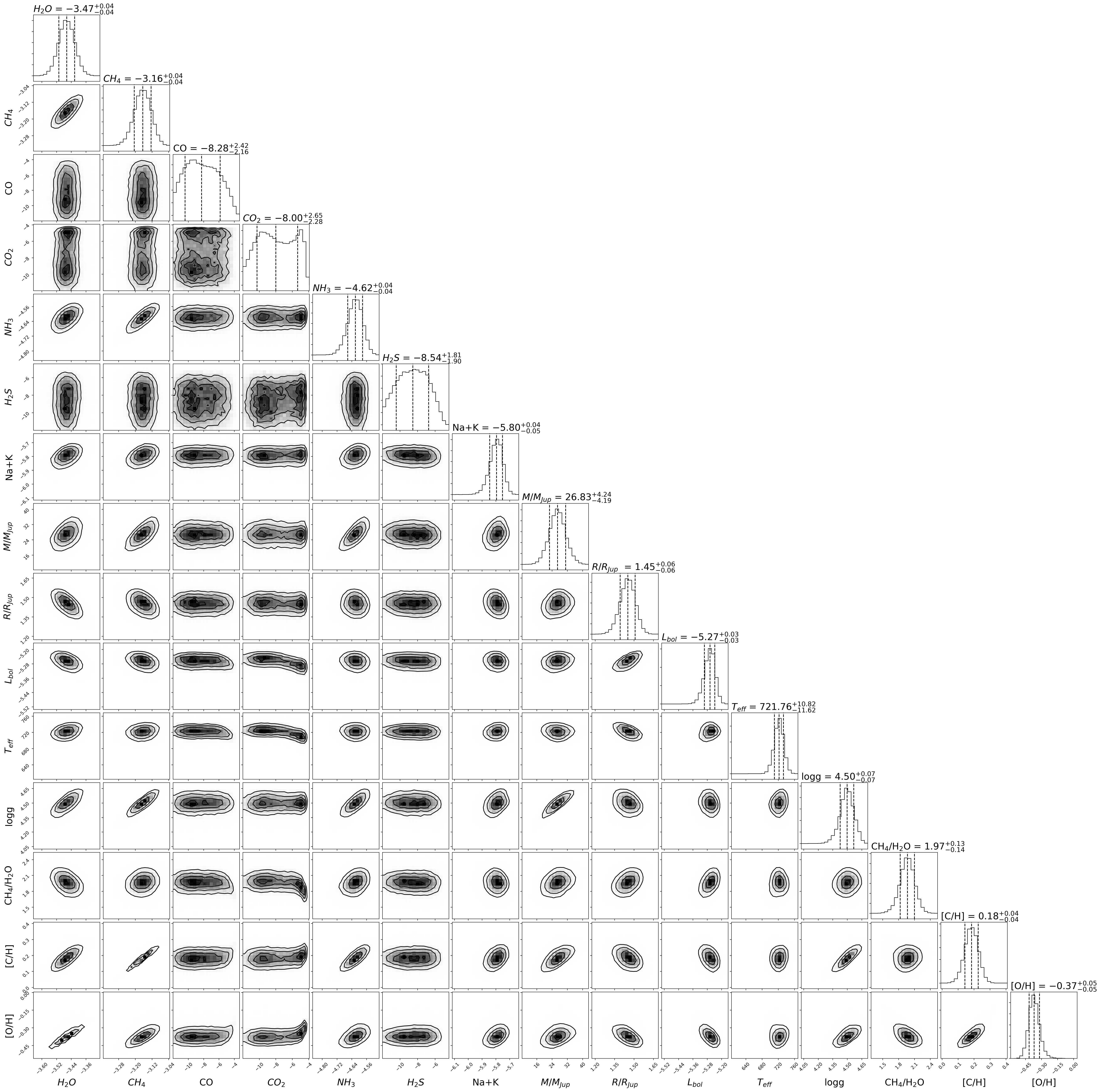}
\caption{Nested sampling corner plot of highest ranked model, the power-law slab cloud, showing posterior probability distributions in 1D histograms with confidence interval. The 2D histogram shows the correlation between the parameters. The corner plot showing the retrieved gas abundances, radius and mass and extrapolated parameters, $\log~g$, \teff, \lbol, C/O, [C/H], [O/H]. See explanation for extrapolated parameters. The gas abundances are shown as $log_{10}(X)$ where X is the gas fraction. [C/H] and [O/H] are relative to solar. 
\label{fig:postcorner}}
\end{figure*}

%%%%%%%%%%%%%%%%Analysis & Discussion%%%%%%%%%%%%%%%%%%%%%%

\section{Analysis and Discussion} \label{sec:analysis}
In line with earlier studies of Ross~458c, we find that the atmosphere is best described by a cloudy model. The best fit cloud model is a power law slab cloud, this is consistent for both the EMCEE and the nested sampling. 

%%%%%%%%%%%%%%%%%%%%%%%%%%%%%%%%%%%%%%%%%%%%%%%%%%%

\subsection{\teff, luminosity, radius and mass}
The bulk properties of mass, radius, \teff\ and luminosity are a mixture of retrieved values, and values extrapolated from the forward model in post processing. The mass and radius are retrieved directly in the PyMultiNest version of Brewster, having been previously inferred from log g and distance scaling in the EMCEE version. The luminosity is found by extrapolating the forward model for the retrieved parameters across a wide wavelength range from 0.5 - 20 \micron, encapsulating essentially all the flux from the target. The \teff\ is then found using the extrapolated \lbol, and the retrieved radius.

% Figure~\ref{fig:sonora} shows a comparison of bulk properties of Ross~458c, together with the Sonora Bobcat evolutionary models \citep{SaumonMarley2021}, for 3 different ages, at 0.1 Gyr, 0.6 Gyr and 1.0 Gyr, which were chosen as likely possible ages for Ross~458c, a somewhat young object. 
 Our extrapolated luminosity for Ross~458c is $-5.27 \pm 0.03$, which is much higher than the value of $\log(L_{bol}/\Lsun) = -5.61 \pm 0.03$ that was found in \citet{Burningham2011}. 
This is may be due to missing absorption from CO and/or \carbondioxide, which  occur in the 4 - 5 \micron\ region, but which are unconstrained in our model, since it is based on the near-infrared only. 

\begin{figure*}
    \centering
    \includegraphics[width=280pt]{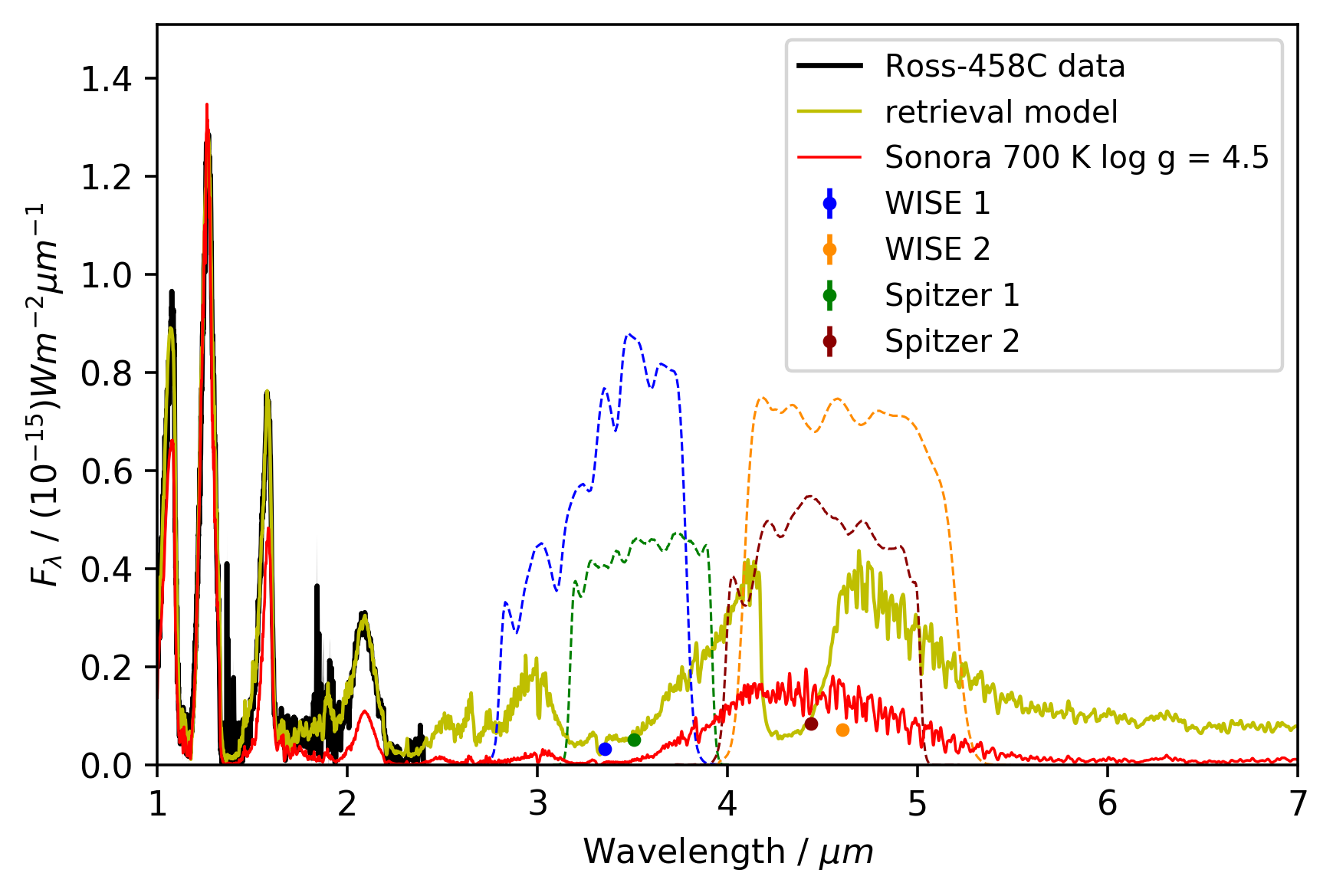}
    \caption{Extrapolated retrieval spectrum of Ross~458c, with the photometry points from WISE1 and 2, and Spitzer 1 and 2 added, as well as transmission regions of the filters, and the 700 K, log g = 4.5 Sonora model.}  
    \label{fig:extrapolated}
\end{figure*}

Figure~\ref{fig:extrapolated} shows that our extrapolated spectrum is a good match to the {\it Spitzer} Channels 1 and 2 photometry which are driven by well-characterised CH$_4$ absorption. The WISE W1 band photometry also matches our model flux reasonably well. However, our extrapolated model is much brighter than the WISE W2 photometry. This filter coincides more strongly with the $4.6~\micron$ CO absorption, and this support our assertion that this is the origin of the mismatch between our extrapolated luminosity and that measured previously by \citet{Burningham2011}. We also note that our extrapolated spectrum shows a strong \carbondioxide\ absorption starting at 4.3 $\mu$m, which is extrapolated from our near-infrared spectrum where its abundance is unconstrained. This is likely spurious and does not warrant further interpretation.

It also plausible that our poorly constrained temperatures at shallower-pressures are also impacting the flux at longer wavelengths. 
Our inferred $T_{\rm eff}$ which is based on this suspect luminosity should be similarly treated with caution. 

We do not think photometric or astrometric uncertainties have a significant effect on our extrapolated luminosity. The astrometric uncertainties are marginalised over in the retrieval model, and so are already incorporated into the radius uncertainty. We have not marginalised over the photometric uncertainty contribution to the flux calibration. However, the photometric uncertainty is at the 1 \% level and thus smaller than our 3 \% uncertainty in luminosity. 

%This value from \citet{Burningham2011}, together with the \teff\ estimate, is shown in pink with error bars on the fourth subplot, with the pink shaded region being the empirical luminosity. 

%\begin{figure*}
   % \centering
   % \includegraphics[width=400pt]{Sonora_ages220822.png}
    %\caption{Comparison of our retrieved and extrapolated parameters with the Sonora Bobcat cloud-free evolutionary models \citep{SaumonMarley2021} for solar metallicity and [M/H] = +0.5 (dashed lines), and three different ages, as indicated on the plot.  The retrieved values are indicated with black error bars. In addition, the lower right figure shows the luminosity and mass as found in %\citet{Burningham2011}, which is the pink error bars, with the shaded pink being the luminosity error.}  
    %\label{fig:sonora}
%\end{figure*}

\begin{figure*}
    \centering
    \includegraphics[width=400pt]{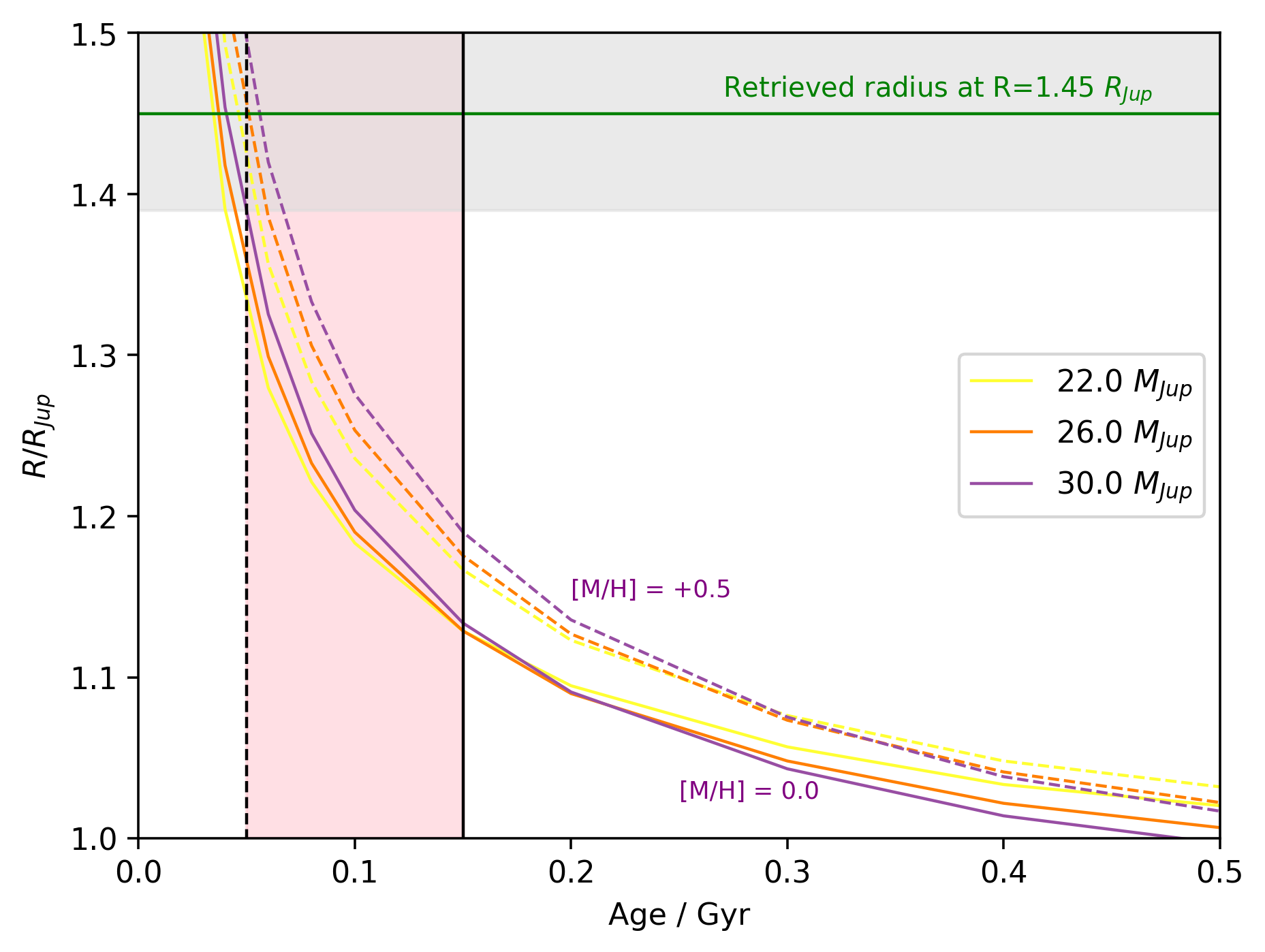}
    \caption{Sonora Bobcat evolutionary models \citep{MarleyNEW2021} of radius against age. The vertical lines and shaded pink area are lower age estimates of Ross 458c. The retrieved radius with shaded errors is shown at R = 1.45 \Rjup. The radius is shown for two different metallicities, of solar and [M/H] = +0.5, and the masses chosen are in the range of our retrieved mass.}  
    \label{fig:evo_age}
\end{figure*}

Figure~\ref{fig:postcorner}, also shows the values for mass and radius, of $27 \pm 4\ \Mjup$ and $1.45 \pm 0.06 \ \Rjup$, respectively. The mass and radius are retrieved directly. 
However, the value for the retrieved radius is driven by the retrieved model scaling factor ($R^2/D^2$) and the Gaia parallax. The retrieved mass value is driven by the influence it has on $\log g$ and the retrieved radius. 

\citet{Burgasser2010} estimate a lower age limit for Ross~458 system of 150~Myr based on the absence of spectroscopic features suggestive of low surface gravity that are seen in young M~dwarfs up to the age of the Pleiades. 
However, this system has a metallicity of ${\rm [M/H]} = +0.2$ (and possibly higher), which is higher than the Pleiades \citep{soderblom2009} , and thus may not follow the same trend in spectroscopic features. 
\citet{Burgasser2010} finds the absence of \ion{Li}{I} absorption at 6708~\AA\ suggests a more robust lower age limit of 30 - 50~Myr, which is consistent with the age of the IC 2391 Moving Group, as found by \citet{Nakajima2010}. 

Figure~\ref{fig:evo_age} shows the radius as a function of age, with two lower limits for the age of 50~Myr and 150~Myr highlighted in \citet{Burgasser2010}. Our retrieved radius of R = 1.45~$R_{Jup}$ coincides with the evolutionary model predictions for ${\rm [M/H]} = +0.5$ and an age near lower of the two limits. The retrieved mass and gravity are similarly consistent with this lower age limit. 

However, if the older age limit of 150~Myr is applied, then both the mass and radius are larger than expected from these models. None-the-less, they result in a value for $\log~g$ which is consistent with previous estimates, for which values are seen in Table~\ref{tab:parameterslitvals}.
We note also, that the EMCEE based retrievals (see Appendix~\ref{sec:app}) find a value for $\log g = 5.13^{+0.20}_{-0.37}$, which is somewhat higher. 
Although its large error, particularly towards lower values mean that it also consistent with previous estimates.

%\citet{Burningham2011} shows the spectrum of Ross~458c with the T8 and T9 standards overplotted, this shows that less than expected \water\ absorption around 1.5 micron is present, compared to the T8 standard, which might be resulting from the high C/O (\meth/\water) ratio. 

%%%%%%%%%%%%%%%%%%%%%%%%%%%%%%%%%%%%%%%%%%%%%%%%%%%%%%

\subsection{C/O ratio}

The C/O ratio can be calculated using the \meth\ to ~\water\ ratio, as CO and \carbondioxide\ are undetected in the NIR. This method for calculating the C/O ratio from methane and water is also used in \citet{Line2015, Line2017} and \citet{Gonzales2020} (the latter with the addition of CO for the L~dwarf). If we naively apply this here, we arrive at an improbably high value of $1.97^{+0.13}_{-0.14}$. This is likely to be outside the scope of any stellar C/O ratio, for which ${\rm C/O > 1}$ is rarely seen. Generally the distribution of stellar C/O is tight, with a peak near the solar value \citep{Nissen2013,Nakajima2016,BrewerFischer2017}.

This section investigates several possibilities to assess their impact on the C/O ratio. We first consider observational and analytical sources of bias, before considering atmospheric origins.

\subsubsection{Observational and model biases}

Telluric water bands are common in the infrared and originate from water absorption in the Earth's atmosphere, which may bias our C/O ratio. We performed a test to check if removing telluric water bands alters our estimate for the \water\ abundance and hence the C/O ratio. This was done by removing all data points in the 1.35 to 1.42 and the 1.80 to 1.95 \micron \ range, where telluric water features are present, and running the nested sampling winning model (power law slab cloud) with the bands removed. Removing the telluric absorption bands did not change the retrieved water abundance significantly, and the \meth/\water\ remained high at 1.88. 

As listed in Tables \ref{tab:emceeresults} and \ref{tab:multinestresults}, the C/O was consistently high across all cloud models tested and sampling methods tested.

%%%%%%%%%%%%%%%%%%%%%%%%%%%%%%%%%%%%%%%%%%%%%%%%%%%%

\subsubsection{Oxygen depletion by condensation}

Examination of our [C/H] and [O/H] values can reveal if the atmosphere is carbon rich or oxygen poor.  The primary stars, Ross~458A and~B are known to have supersolar metallicities of ${\rm [Fe/H]} = +0.2$ \citep{Burgasser2010}, which agree with the retrieved ${\rm [C/H]} = +0.18$ of Ross 458c, meaning that the carbon abundance is consistent with the primary. However, Ross 458c's ${\rm [O/H]} = -0.37 \pm 0.05$, suggesting it is oxygen poor or otherwise depleted.

A certain amount of oxygen can be tied up via silicate condensation and would be “missing” from the atmosphere. However, under typical assumptions for T~dwarf atmospheres this would not account for all of the missing oxygen. By allowing 25\% of oxygen to be tied up in silicate clouds \citep{BurrowsSharp1999}, the C/O ratio would lower to 1.47, which is still extremely high.

Water clouds in the atmosphere of Ross~458c would account for the missing water and lower the C/O ratio substantially, but the temperature is too high for water clouds. Water clouds are expected to become a significant opacity source in brown dwarfs of temperatures less than 400 K \citep{Morley2014}. 

\subsubsection{Non-equilibrium chemistry} 
\label{sec:noneq}
The value of the \meth/\water\ ratio is likely impacted by non-equilibrium chemistry which will lead to more oxygen being tied up in CO and CO$_2$ than would otherwise be expected at the low-temperatures of late-T dwarfs. Previous work has highlighted the importance of non-equilibrium chemistry, leading to higher CO and \carbondioxide\ abundances.

In the cooler T dwarfs, convection drives the disequilibrium chemistry. \citet{Miles2020} find that disequilibrium chemistry plays an important role in directly imaged exoplanets and brown dwarfs, leading to important CO absorption in the spectra.

\citet{Noll1997} found that for Gliese 229B, CO was also present at larger abundances than expected, such as in the 4.7 micron band, as a disequilibrium species high up in the atmosphere, suggesting transport induced quenching. 

CO is favoured as the dominant C-bearing gas at high temperatures (deep in the atmosphere), whereas \meth\ is favoured as the dominant C-bearing gas at low temperatures \citep[higher altitudes, ][]{LoddersFegley2002, Visscher2012}.
Through atmospheric mixing, the chemical composition can be driven out of equilibrium. Rapid vertical mixing can transport a gas to higher altitude (and lower temperatures) before the chemical constituents have had time to reach chemical equilibrium. If the mixing time is faster than the chemical reaction time, disequilibrium can occur. 

At the quench point, (where the chemical timescale = mixing timescale), the abundance of the species is ``quenched" at a fixed value as it is mixed to higher altitudes. 
CO + CO$_{2}$ are subject to transport induced quenching, which may lead to a quenched abundance in the upper atmosphere far in excess of that predicted by equilibrium, if undergoing rapid vertical mixing \citep{Visscher2010}.

Figure \ref{fig:NEWabundance} shows a good match with thermochemical predictions at high pressure and temperature ($T \approx 2000$K). This suggests that the observed \meth/\water\ ratio can be consistent with C/O = 1.35 (the maximum modelled value in our thermochemical grid), if carbon-oxygen chemistry is quenched at the 100~bar, 2000~K level. However, at 2000~K the chemical reactions converting CO/\meth\ will be fast, suggesting that it is unlikely that the observed abundances can be attributed to quench chemistry alone \citep{VisscherMoses2011}.

\begin{figure}
    \includegraphics[width=250pt]{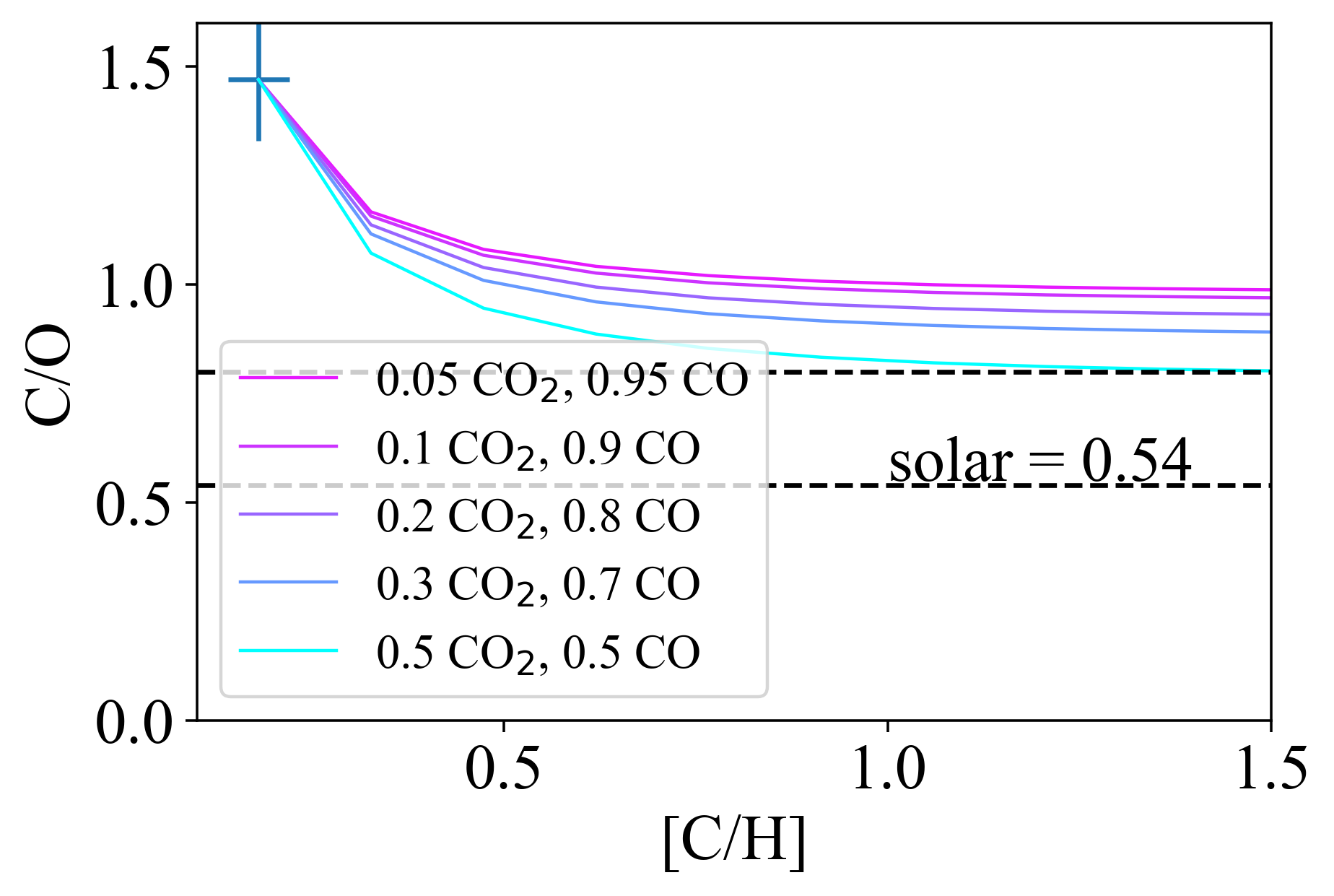}
    \caption{The impact on the implied C/O ratio of incorporating CO and CO$_2$ that may have been missed by our NIR spectrum. The base-case indicated by the point with error bars at the top-left of the plot is where we estimate C/O assuming all carbon and oxygen (after depletion by condensing clouds) exists in the \water\ and \meth\ that we have measured. }The lines indicate how the C/O changes as we allow for increasing quantities of CO and CO$_2$ at the proportions indicated and tracked by the corresponding increase in [C/H]. The top dashed line indicates C/O = 0.8, the upper range seen in solar type stars, and the \citet{Asplund2009} Solar value is indicated by the lower line. 
    \label{fig:COsim}
\end{figure}

We have simulated how the estimated C/O ratio can be affected by incorporating CO and CO$_2$ that might have been missed by our near-infrared spectroscopy. In Figure~\ref{fig:COsim} we show how the C/O ratio decreases as more CO and CO$_2$ are incorporated,  as indicated via the associated increase in the [C/H] metallicity. 
The plot is based on our condensation-corrected C/O$ = 1.47$, based on \meth/\water\ alone, which corresponds to  ${\rm [C/H]} = +0.18$ (see Figure~\ref{fig:postcorner}).
As expected, adding CO drags the C/O ratio closer to 1.  However, a substantial (0.5) fraction of CO$_2$ is required to bring the C/O ratio to within the typical stellar range (i.e.
$\leq \sim 0.8$, see Section~\ref{sec:COstars}). 
 
As discussed in \citet{Burningham2011}, there are disagreements as to the full network of CO-CO$_2$-CH$_4$ reactions, and where their equilibria lie.  But, even the most pro-CO$_2$ cases in the BT Settl model grids discussed in that paper have CO outnumbering CO$_2$ by a factor of 20, which is the most CO$_2$ {\it poor} case illustrated in Figure~\ref{fig:COsim}. 

In addition to the high abundance of CO$_2$ required to bring the implied C/O ratio towards the expected stellar range, it is also clear that the implied [C/H] must rise significantly also. In all cases, allowing for enough CO and CO$_2$ to bring C/O below 1.0 implies ${\rm [C/H]} > +0.5$, which is significantly higher than suggested by the metallicty of the primary stars in the system, and beyond the range of carbon abundances or metallicities seen in the solar neighbourhood \citep[e.g.][]{Nissen2013,hinkel2014}. 

In the absence of spectroscopy covering the missing CO and CO$_2$ absorption, we argue that the estimate of the C/O ratio made via comparison of our retrieved abundances with our thermochemical model grids (Figure~\ref{fig:NEWabundance}) as the most reasonable, i.e. C/O$\approx 1.35$. We note that this value implicitly includes the impact of rain-out of oxygen in condensates according to phase-equilibrium chemistry at solar abundance ratios.

\subsubsection{C/O of primary system}
\label{sec:COstars}
The C/O ratio of the primary system, Ross~458AB, is unknown, and it may not share the solar value of 0.54 \citep{Asplund2005}.  
However, the range of stellar C/O ratios for solar type stars is relatively narrow, with C/O $> 0.8$ very rare. As the primaries are not carbon stars, a C/O $> 0.8$ would be highly unlikely. The peak of the distribution is around 0.47, with a tail towards lower and higher C/O ratios \citep{Nissen2013,BrewerFischer2017}. This makes the high implied value for Ross~458c particularly interesting. 
%%%%%%%%%%%%%%%%%%%%%%%%%%%%%%%%%%%%%%%%%%%%%%%%%%

\subsection{Interpretation of C/O ratio}

Taking the above considerations into account, the fact that the C/O ratio appears to be so different from its primaries, may suggest a planetary formation route for Ross~458c. This formation would have to be such that it is not enriched with oxygen. Ross~458c has not necessarily formed in situ but could have migrated to its current position. The protoplanetary disk would have different compositions at different radii, at the various ice lines. In order for the atmosphere to be oxygen poor, the formation must have taken place outside of the \water \ snowline, with a process that inhibits accretion of oxygen bearing silicates and allows for more carbon-rich gas to accrete. 
The suggestion of a planetary formation route for Ross~458c seems unlikely due to its mass ratio with the primaries and our own large retrieved mass (driven up by the large retrieved radius). 

This oxygen depletion that we are seeing, is also noted in other retrievals, such as \citet{Calamari2022}, who find a supersolar C/O ratio for Gliese 229B. 
They compare their work to other works done on T-dwarfs, who likewise see the trend of supersolar C/O \citep{Line2017, Zalesky2019, Zalesky2022}.
They speculate that unanticipated chemistry may provide additional oxygen sinks, such as unmodeled magnesium silicates and/or iron-bearing condensates, that could drive an apparently super-solar C/O.

%However, given the substantial disagreement of our mass and radius with theoretical expectations, perhaps both should be viewed as suspect. 

\section{Conclusion}

In this work we have presented the first retrieval of Ross~458c. Ross~458c is best parameterised by a non-grey (power-law) cloud, structured as a slab. 
Both the radius and the mass are overestimated, based on evolutionary models. The bolometric luminosity is higher than found in \citet{Burningham2011}.

The \meth/\water\ ratio is much higher than expected, at 1.97. This is due to missing oxygen from a low water abundance. Comparisons to thermochemical grid models suggest a still high C/O ratio of 1.35, if \meth\ and \water\ is quenched at 2000 K due to vigorous mixing. 

The ${\rm [C/H]} = +0.18$ matches up with the metallicity of Ross~458AB of [Fe/H] = +0.2 \citep{Burgasser2010}. 

Even with oxygen sequestered into clouds and accounting for transport induced quenching, the C/O ratio remains high.  
This points to either a planetary origin of Ross~458c, or the presence on an unidentified sink for the gas-phase oxygen.

These retrieval results are the first in an analysis of a larger sample of late-T dwarfs that seek to answer the following questions: Do companion T dwarfs have different composition than their host star? Is the C/O ratio of free-floating T dwarfs different to that of the stellar population and to that of companion T dwarfs? What trends in C/O ratio is present across the T dwarf range? 

\subsubsection{Future work}
In order to compare the C/O ratios of companions to their primary star(s), we need the abundances of the primary. This could be done following the method of \citet{TsujiNakajima2014,TsujiNakajima2016}, which uses high resolution spectra of \water\ and CO in the near-infrared bands.

Additionally, by extending the spectrum of Ross~458c to 5 \micron, we would include CO and \carbondioxide\ absorption which might be a significant contributor to carbon and oxygen abundances, altering the C/O ratio. No spectroscopic data at those wavelengths exists yet. 
The data for Ross~458c which will be obtained through JWST at near-infrared and mid-infrared wavelengths and at greater sensitivity will aid our understanding of absorbers in the atmosphere, such as CO and \carbondioxide, and the thermal profile over wider range of pressures. 

\section{Data Availability}
All data underlying this article are publicly available from the relevant
observatory archive, or upon reasonable request to the author.

\section{Acknowledgements}
J.G. acknowledges support from UK
Research and Innovation-Science and Technology Facilities Council (UKRI-STFC) studentships. This research was made possible thanks to the Royal Society International Exchange grant No. IES/R3/170266. E.G. acknowledges support from the Heising-Simons Foundation
for this research. This work benefited from the 2022 Exoplanet Summer Program in the Other Worlds Laboratory (OWL) at the University of California, Santa Cruz, a program funded by
the Heising-Simons Foundation. This work has made use of the University of Hertfordshire’s high-performance computing facility. J.F acknowledges the NSF award 1909776 and NASA XRP Award 80NSSC22K0142. R.L acknowledges the following grants: JWST-AR-01977.007-A, JWST-AR-02232.008A, 80NSSC22K0953 (NASA ROSES XRP).

\appendix

\section{Comparison of EMCEE and Pymultinest results}
\label{sec:app}
Table~\ref{tab:comparison} shows the comparison of retrieved parameters between the EMCEE and the PyMultiNest sampler. The values are consistent with each other to 1 $\sigma$ for the most part, with a couple of exceptions that differ by under 2 $\sigma$.

\begin{center} 
\begin{table*} 
\begin{tabular}{c c c}
\hline\hline
     Parameter & EMCEE & PyMultiNest  \\
     \hline 
     \water & $-3.00^{+0.16}_{-0.24}$ & $-3.47^{+0.04}_{-0.04}$\\
     \meth & $-2.76^{+0.14}_{-0.30}$ & $-3.16^{+0.04}_{-0.04}$ \\
     CO & $-7.81^{+2.73}_{-2.83}$ & $-8.28^{+2.42}_{-2.16}$ \\
     \carbondioxide & $-7.36^{+3.07}_{-3.87}$ & $-8.00^{+2.65}_{-2.28}$ \\
     $NH_{3}$ & $-4.38^{+0.20}_{-2.01}$ & $-4.62^{+0.04}_{-0.04}$ \\
     $H_{2}S$ & $-8.82^{+2.63}_{-2.11}$ & $-8.54^{1.81}_{1.90}$\\
     Na+K & $-6.12^{+0.37}_{-1.31}$ & $-5.80^{+0.04}_{-0.05}$ \\
     log g & $5.13^{+0.20}_{-0.37}$ & $4.50^{+0.07}_{-0.07}$\\
     \hline 
\end{tabular}
\caption{Comparison of parameters between EMCEE and PyMultiNest for the power law slab cloud (the winning model from PyMultiNest, and the second ranked model from EMCEE, in order to allow for a more direct comparison. 
\label{tab:comparison}}
\end{table*}
\end{center}

\begin{figure*}
    \centering
    \includegraphics[width=280pt]{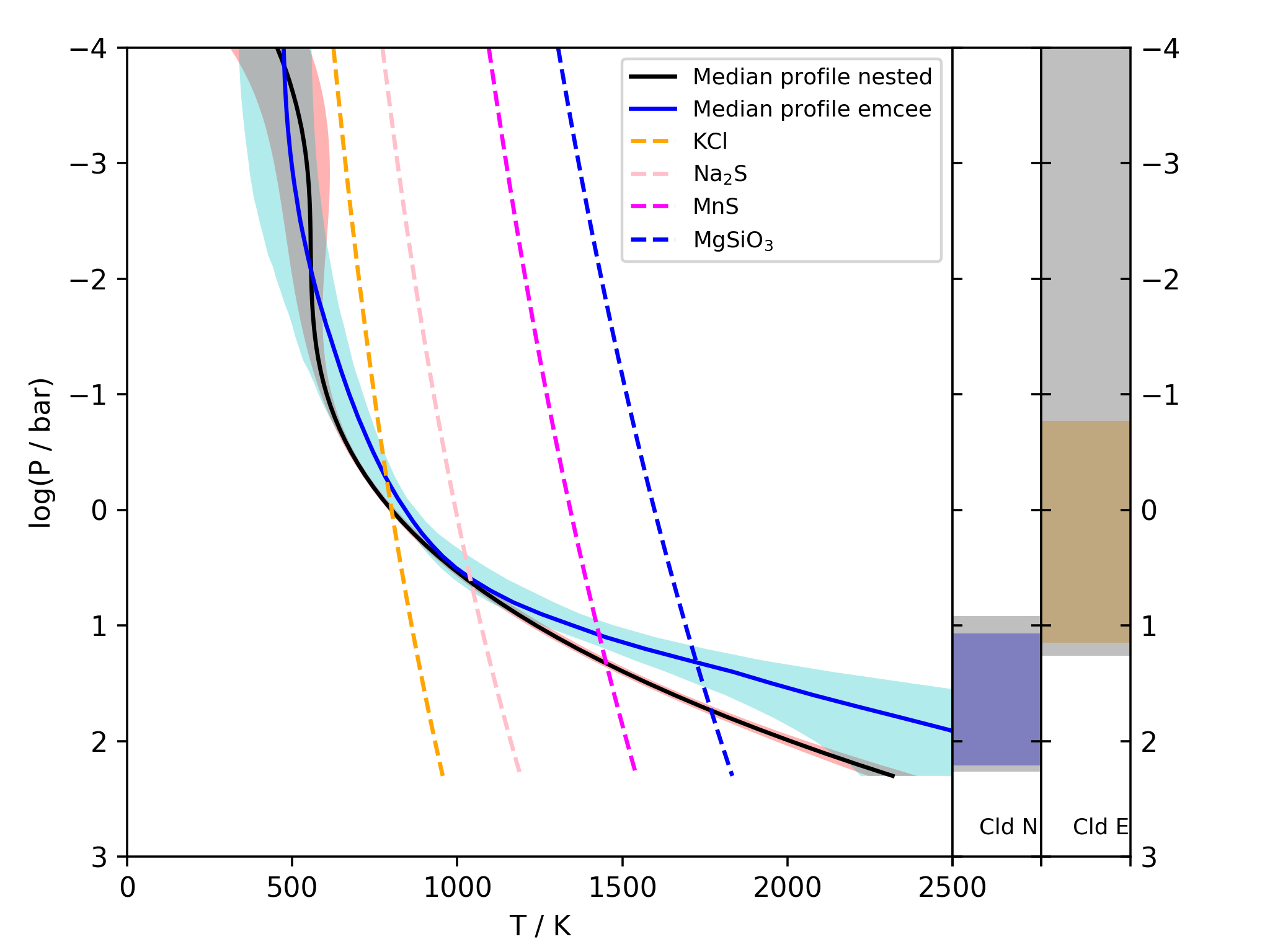}
    \caption{Comparison of the thermal profiles with their errors, for the power-law slab cloud of the EMCEE (in blue) and PyMultiNest (in black) samplers with the cloud condensation curves overplotted. Also plotted are the cloud locations in the bars on the rights (cloud N being PyMultiNest and cloud E being the EMCEE sampler), with their associated errors in grey.}  
    \label{fig:nest_emcee_sameplot}
\end{figure*}

Figure~\ref{fig:nest_emcee_sameplot} shows the comparison of the thermal profiles of the EMCEE and PyMultiNest sampler for the power-law slab cloud. They follow similar profiles, with the EMCEE profile being slightly warmer at deeper pressures.

The location of the EMCEE cloud is unconstrained within the atmosphere at the top of the cloud. 

\begin{center} 
\begin{table*}
\begin{tabular}{c c c}
\hline\hline
     Model type & $\textrm \meth/\water$ PyMultiNest & $\textrm \meth/\water$ EMCEE \\
     \hline 
     Power law slab cloud &   $1.97^{+0.13}_{-0.14}$ & $1.69^{+0.31}_{-0.51}$\\
     Power law slab cloud, Allard alkalis & $2.13^{+0.14}_{-0.15}$ & $1.82^{+0.31}_{-0.46}$\\
     \enstatite\ slab cloud & $1.86^{+0.19}_{-0.18}$ & $1.42^{+0.37}_{-0.42}$ \\
     \sodiumsulphide\ slab cloud &  $1.82^{+0.20}_{-0.19}$ & $1.83^{+0.19}_{-0.23}$ \\
     Cloud free &  $1.74^{+0.20}_{-0.19}$ & $1.83^{+0.19}_{-0.23}$ \\
     Grey deck cloud & $1.68^{+0.20}_{-0.20}$ & $1.57^{+0.32}_{-0.57}$ \\
     KCl deck cloud &  $1.79^{+0.17}_{-0.18}$ & $1.71^{+0.27}_{-0.31}$\\
     \enstatite\ deck cloud &  $1.76^{+0.19}_{-0.18}$ & $1.71^{+0.41}_{-1.06}$ \\
     Power law deck cloud &  $1.70^{+0.20}_{-0.21}$ & $1.61^{+0.36}_{-0.80}$ \\
     Soot deck cloud &  $1.32^{+0.28}_{-0.23}$ & $1.52^{+0.26}_{-0.23}$ \\
     KCl slab cloud &  $1.73^{+0.18}_{-0.17}$ & $1.28^{+0.32}_{-0.21}$ \\
     Grey slab cloud &  $1.63^{+0.17}_{-0.16}$ & $1.67^{+0.28}_{-0.56}$ \\
     MnS slab cloud &  $1.71^{+0.17}_{-0.17}$ & $1.71^{+0.24}_{-0.29}$ \\
     MnS deck cloud &  $1.75^{+0.19}_{-0.21}$ & $1.66^{+0.32}_{-0.64}$ \\
     Soot slab cloud & $1.45^{+0.23}_{-0.37}$ & $1.79^{+0.22}_{-0.24}$ \\
     %\sodiumsulphide deck cloud & 20 & 1.26 & -21.34\\
     \hline 
\end{tabular}
\caption{Comparison of the \meth/\water\ ratio between PyMultiNest and EMCEE, for the different models tested. 
\label{tab:co_comparison} }
\end{table*}
\end{center}

Table~\ref{tab:co_comparison} shows the comparison of $\textrm \meth/\water$ for the different models tested, for both the PyMultinest and the EMCEE sampler. The $\textrm \meth/\water$ stays consistently high across all models, for both sampling methods. 

The two highest ranking models are the same for both samplers, just with the order switched. See Table~\ref{tab:multinestresults} and \ref{tab:emceeresults} for model rankings for both samplers.

%\begin{figure*}
    %\centering
    %\includegraphics[width=280pt]{emcee_corner_ross458c_89slab.pngg}
    %\caption{Corner plot for EMCEE second ranked model, the power law slab cloud, which was the winning model for nested sampling.}  
    %\label{fig:emcee_corner}
%\end{figure*}

%Figure~\ref{fig:emcee_corner} shows the corner plot for the EMCEE sampler, with the power law slab cloud. 

\bibliographystyle{mnras}
\bibliography{Bibliography}

\begin{thebibliography}{}
\makeatletter
\relax
\def\mn@urlcharsother{\let\do\@makeother \do\$\do\&\do\#\do\^\do\_\do\%\do\~}
\def\mn@doi{\begingroup\mn@urlcharsother \@ifnextchar [ {\mn@doi@}
  {\mn@doi@[]}}
\def\mn@doi@[#1]#2{\def\@tempa{#1}\ifx\@tempa\@empty \href
  {http://dx.doi.org/#2} {doi:#2}\else \href {http://dx.doi.org/#2} {#1}\fi
  \endgroup}
\def\mn@eprint#1#2{\mn@eprint@#1:#2::\@nil}
\def\mn@eprint@arXiv#1{\href {http://arxiv.org/abs/#1} {{\tt arXiv:#1}}}
\def\mn@eprint@dblp#1{\href {http://dblp.uni-trier.de/rec/bibtex/#1.xml}
  {dblp:#1}}
\def\mn@eprint@#1:#2:#3:#4\@nil{\def\@tempa {#1}\def\@tempb {#2}\def\@tempc
  {#3}\ifx \@tempc \@empty \let \@tempc \@tempb \let \@tempb \@tempa \fi \ifx
  \@tempb \@empty \def\@tempb {arXiv}\fi \@ifundefined
  {mn@eprint@\@tempb}{\@tempb:\@tempc}{\expandafter \expandafter \csname
  mn@eprint@\@tempb\endcsname \expandafter{\@tempc}}}

\bibitem[\protect\citeauthoryear{{Allard}, {Kielkopf}  \& {Allard}}{{Allard}
  et~al.}{2007a}]{Allard2007a}
{Allard} N.~F.,  {Kielkopf} J.~F.,   {Allard} F.,  2007a, \mn@doi [European
  Physical Journal D] {10.1140/epjd/e2007-00230-6}, \href
  {https://ui.adsabs.harvard.edu/abs/2007EPJD...44..507A} {44, 507}

\bibitem[\protect\citeauthoryear{{Allard}, {Spiegelman}  \&
  {Kielkopf}}{{Allard} et~al.}{2007b}]{Allard2007b}
{Allard} N.~F.,  {Spiegelman} F.,   {Kielkopf} J.~F.,  2007b, \mn@doi [\aap]
  {10.1051/0004-6361:20066616}, \href
  {https://ui.adsabs.harvard.edu/abs/2007A&A...465.1085A} {465, 1085}

\bibitem[\protect\citeauthoryear{{Allard}, {Spiegelman}  \&
  {Kielkopf}}{{Allard} et~al.}{2016}]{Allard2016}
{Allard} N.~F.,  {Spiegelman} F.,   {Kielkopf} J.~F.,  2016, \mn@doi [\aap]
  {10.1051/0004-6361/201628270}, \href
  {https://ui.adsabs.harvard.edu/abs/2016A&A...589A..21A} {589, A21}

\bibitem[\protect\citeauthoryear{{Asplund}, {Grevesse}  \& {Sauval}}{{Asplund}
  et~al.}{2005}]{Asplund2005}
{Asplund} M.,  {Grevesse} N.,   {Sauval} A.~J.,  2005, in {Barnes} Thomas~G.
  I.,  {Bash} F.~N.,  eds,  Astronomical Society of the Pacific Conference
  Series Vol. 336, Cosmic Abundances as Records of Stellar Evolution and
  Nucleosynthesis. p.~25

\bibitem[\protect\citeauthoryear{{Asplund}, {Grevesse}, {Sauval}  \&
  {Scott}}{{Asplund} et~al.}{2009}]{Asplund2009}
{Asplund} M.,  {Grevesse} N.,  {Sauval} A.~J.,   {Scott} P.,  2009, \mn@doi
  [\araa] {10.1146/annurev.astro.46.060407.145222}, \href
  {https://ui.adsabs.harvard.edu/abs/2009ARA&A..47..481A} {47, 481}

\bibitem[\protect\citeauthoryear{{Bakos} et~al.,}{{Bakos}
  et~al.}{2009}]{bakos2009}
{Bakos} G.~{\'A}.,  et~al., 2009, \mn@doi [\apj] {10.1088/0004-637X/707/1/446},
  \href {https://ui.adsabs.harvard.edu/abs/2009ApJ...707..446B} {707, 446}

\bibitem[\protect\citeauthoryear{{Bate}, {Bonnell}  \& {Bromm}}{{Bate}
  et~al.}{2002}]{BateBonnellBromm2002}
{Bate} M.~R.,  {Bonnell} I.~A.,   {Bromm} V.,  2002, \mn@doi [\mnras]
  {10.1046/j.1365-8711.2002.05539.x}, \href
  {https://ui.adsabs.harvard.edu/abs/2002MNRAS.332L..65B} {332, L65}

\bibitem[\protect\citeauthoryear{{Bell}}{{Bell}}{1980}]{Bell1980}
{Bell} K.~L.,  1980, Journal of Physics B Atomic Molecular Physics, 13, 1859

\bibitem[\protect\citeauthoryear{{Bell} \& {Berrington}}{{Bell} \&
  {Berrington}}{1987}]{Bell1987}
{Bell} K.~L.,  {Berrington} K.~A.,  1987, Journal of Physics B Atomic Molecular
  Physics, 20, 801

\bibitem[\protect\citeauthoryear{{Beuzit} et~al.,}{{Beuzit}
  et~al.}{2004}]{Beuzit2004}
{Beuzit} J.~L.,  et~al., 2004, \mn@doi [\aap] {10.1051/0004-6361:20048006},
  \href {https://ui.adsabs.harvard.edu/abs/2004A&A...425..997B} {425, 997}

\bibitem[\protect\citeauthoryear{{Bouy} et~al.,}{{Bouy}
  et~al.}{2022}]{bouy2022}
{Bouy} H.,  et~al., 2022, arXiv e-prints, \href
  {https://ui.adsabs.harvard.edu/abs/2022arXiv220600916B} {p. arXiv:2206.00916}

\bibitem[\protect\citeauthoryear{{Brewer} \& {Fischer}}{{Brewer} \&
  {Fischer}}{2017}]{BrewerFischer2017}
{Brewer} J.~M.,  {Fischer} D.~A.,  2017, \mn@doi [\apj]
  {10.3847/1538-4357/aa6d53}, \href
  {https://ui.adsabs.harvard.edu/abs/2017ApJ...840..121B} {840, 121}

\bibitem[\protect\citeauthoryear{{Buchner} et~al.,}{{Buchner}
  et~al.}{2014}]{Buchner2014}
{Buchner} J.,  et~al., 2014, \mn@doi [\aap] {10.1051/0004-6361/201322971},
  \href {https://ui.adsabs.harvard.edu/abs/2014A&A...564A.125B} {564, A125}

\bibitem[\protect\citeauthoryear{{Burgasser}, {Geballe}, {Golimowski},
  {Leggett}, {Kirkpatrick}, {Knapp}  \& {Fan}}{{Burgasser}
  et~al.}{2003}]{Burgasser2003}
{Burgasser} A.~J.,  {Geballe} T.~R.,  {Golimowski} D.~A.,  {Leggett} S.~K.,
  {Kirkpatrick} J.~D.,  {Knapp} G.~R.,   {Fan} X.,  2003, in {Mart{\'\i}n} E.,
  ed.,  IAU Symposium Vol. 211, Brown Dwarfs. p.~377

\bibitem[\protect\citeauthoryear{{Burgasser} et~al.,}{{Burgasser}
  et~al.}{2010}]{Burgasser2010}
{Burgasser} A.~J.,  et~al., 2010, \mn@doi [\apj]
  {10.1088/0004-637X/725/2/1405}, \href
  {https://ui.adsabs.harvard.edu/abs/2010ApJ...725.1405B} {725, 1405}

\bibitem[\protect\citeauthoryear{{Burningham} et~al.,}{{Burningham}
  et~al.}{2011}]{Burningham2011}
{Burningham} B.,  et~al., 2011, \mn@doi [\mnras]
  {10.1111/j.1365-2966.2011.18664.x}, \href
  {https://ui.adsabs.harvard.edu/abs/2011MNRAS.414.3590B} {414, 3590}

\bibitem[\protect\citeauthoryear{{Burningham}, {Marley}, {Line}, {Lupu},
  {Visscher}, {Morley}, {Saumon}  \& {Freedman}}{{Burningham}
  et~al.}{2017}]{burningham2017}
{Burningham} B.,  {Marley} M.~S.,  {Line} M.~R.,  {Lupu} R.,  {Visscher} C.,
  {Morley} C.~V.,  {Saumon} D.,   {Freedman} R.,  2017, \mn@doi [\mnras]
  {10.1093/mnras/stx1246}, \href
  {https://ui.adsabs.harvard.edu/abs/2017MNRAS.470.1177B} {470, 1177}

\bibitem[\protect\citeauthoryear{{Burningham} et~al.,}{{Burningham}
  et~al.}{2021}]{Burningham2021}
{Burningham} B.,  et~al., 2021, \mn@doi [\mnras] {10.1093/mnras/stab1361},
  \href {https://ui.adsabs.harvard.edu/abs/2021MNRAS.tmp.1380B} {}

\bibitem[\protect\citeauthoryear{{Burrows} \& {Sharp}}{{Burrows} \&
  {Sharp}}{1999}]{BurrowsSharp1999}
{Burrows} A.,  {Sharp} C.~M.,  1999, \mn@doi [\apj] {10.1086/306811}, \href
  {https://ui.adsabs.harvard.edu/abs/1999ApJ...512..843B} {512, 843}

\bibitem[\protect\citeauthoryear{{Burrows} \& {Volobuyev}}{{Burrows} \&
  {Volobuyev}}{2003}]{BurrowsVolobuyev2003}
{Burrows} A.,  {Volobuyev} M.,  2003, \mn@doi [\apj] {10.1086/345412}, \href
  {https://ui.adsabs.harvard.edu/abs/2003ApJ...583..985B} {583, 985}

\bibitem[\protect\citeauthoryear{{Calamari} et~al.,}{{Calamari}
  et~al.}{2022}]{Calamari2022}
{Calamari} E.,  et~al., 2022, \mn@doi [\apj] {10.3847/1538-4357/ac9cc9}, \href
  {https://ui.adsabs.harvard.edu/abs/2022ApJ...940..164C} {940, 164}

\bibitem[\protect\citeauthoryear{Dalzell \& Sarofim}{Dalzell \&
  Sarofim}{1969}]{Dalzell1969}
Dalzell W.~H.,  Sarofim A.~F.,  1969, Journal of Heat Transfer-transactions of
  The Asme, 91, 100

\bibitem[\protect\citeauthoryear{{Eggen}}{{Eggen}}{1960}]{Eggen1960}
{Eggen} O.~J.,  1960, \mn@doi [\mnras] {10.1093/mnras/120.6.540}, \href
  {https://ui.adsabs.harvard.edu/abs/1960MNRAS.120..540E} {120, 540}

\bibitem[\protect\citeauthoryear{{Faherty}, {Rice}, {Cruz}, {Mamajek}  \&
  {N{\'u}{\~n}ez}}{{Faherty} et~al.}{2013}]{faherty2013}
{Faherty} J.~K.,  {Rice} E.~L.,  {Cruz} K.~L.,  {Mamajek} E.~E.,
  {N{\'u}{\~n}ez} A.,  2013, \mn@doi [\aj] {10.1088/0004-6256/145/1/2}, \href
  {https://ui.adsabs.harvard.edu/abs/2013AJ....145....2F} {145, 2}

\bibitem[\protect\citeauthoryear{{Faherty} et~al.,}{{Faherty}
  et~al.}{2021}]{faherty2021}
{Faherty} J.~K.,  et~al., 2021, \mn@doi [\apj] {10.3847/1538-4357/ac2499},
  \href {https://ui.adsabs.harvard.edu/abs/2021ApJ...923...48F} {923, 48}

\bibitem[\protect\citeauthoryear{{Feroz}, {Hobson}, {Cameron}  \&
  {Pettitt}}{{Feroz} et~al.}{2019}]{Feroz2019}
{Feroz} F.,  {Hobson} M.~P.,  {Cameron} E.,   {Pettitt} A.~N.,  2019, \mn@doi
  [The Open Journal of Astrophysics] {10.21105/astro.1306.2144}, \href
  {https://ui.adsabs.harvard.edu/abs/2019OJAp....2E..10F} {2, 10}

\bibitem[\protect\citeauthoryear{{Filippazzo}, {Rice}, {Faherty}, {Cruz}, {Van
  Gordon}  \& {Looper}}{{Filippazzo} et~al.}{2015}]{Filippazzo2015}
{Filippazzo} J.~C.,  {Rice} E.~L.,  {Faherty} J.,  {Cruz} K.~L.,  {Van Gordon}
  M.~M.,   {Looper} D.~L.,  2015, \mn@doi [\apj] {10.1088/0004-637X/810/2/158},
  \href {https://ui.adsabs.harvard.edu/abs/2015ApJ...810..158F} {810, 158}

\bibitem[\protect\citeauthoryear{{Foreman-Mackey}, {Hogg}, {Lang}  \&
  {Goodman}}{{Foreman-Mackey} et~al.}{2013}]{ForemanMackie2013}
{Foreman-Mackey} D.,  {Hogg} D.~W.,  {Lang} D.,   {Goodman} J.,  2013, \mn@doi
  [\pasp] {10.1086/670067}, \href
  {https://ui.adsabs.harvard.edu/abs/2013PASP..125..306F} {125, 306}

\bibitem[\protect\citeauthoryear{{Fortney}}{{Fortney}}{2012}]{Fortney2012}
{Fortney} J.~J.,  2012, \mn@doi [\apjl] {10.1088/2041-8205/747/2/L27}, \href
  {https://ui.adsabs.harvard.edu/abs/2012ApJ...747L..27F} {747, L27}

\bibitem[\protect\citeauthoryear{{Freedman}, {Marley}  \& {Lodders}}{{Freedman}
  et~al.}{2008}]{Freedman2008}
{Freedman} R.~S.,  {Marley} M.~S.,   {Lodders} K.,  2008, \mn@doi [\apjs]
  {10.1086/521793}, \href
  {https://ui.adsabs.harvard.edu/abs/2008ApJS..174..504F} {174, 504}

\bibitem[\protect\citeauthoryear{{Freedman}, {Lustig-Yaeger}, {Fortney},
  {Lupu}, {Marley}  \& {Lodders}}{{Freedman} et~al.}{2014}]{Freedman2014}
{Freedman} R.~S.,  {Lustig-Yaeger} J.,  {Fortney} J.~J.,  {Lupu} R.~E.,
  {Marley} M.~S.,   {Lodders} K.,  2014, \mn@doi [\apjs]
  {10.1088/0067-0049/214/2/25}, \href
  {https://ui.adsabs.harvard.edu/abs/2014ApJS..214...25F} {214, 25}

\bibitem[\protect\citeauthoryear{{Gaia Collaboration} et~al.,}{{Gaia
  Collaboration} et~al.}{2018}]{Gaia}
{Gaia Collaboration} et~al., 2018, \mn@doi [\aap]
  {10.1051/0004-6361/201833051}, \href
  {https://ui.adsabs.harvard.edu/abs/2018A&A...616A...1G} {616, A1}

\bibitem[\protect\citeauthoryear{{Goldman}, {Marsat}, {Henning}, {Clemens}  \&
  {Greiner}}{{Goldman} et~al.}{2010}]{Goldman2010}
{Goldman} B.,  {Marsat} S.,  {Henning} T.,  {Clemens} C.,   {Greiner} J.,
  2010, \mn@doi [\mnras] {10.1111/j.1365-2966.2010.16524.x}, \href
  {https://ui.adsabs.harvard.edu/abs/2010MNRAS.405.1140G} {405, 1140}

\bibitem[\protect\citeauthoryear{{Gonzales}, {Burningham}, {Faherty}, {Cleary},
  {Visscher}, {Marley}, {Lupu}  \& {Freedman}}{{Gonzales}
  et~al.}{2020}]{Gonzales2020}
{Gonzales} E.,  {Burningham} B.,  {Faherty} J.,  {Cleary} C.,  {Visscher} C.,
  {Marley} M.,  {Lupu} R.,   {Freedman} R.,  2020, arXiv e-prints, \href
  {https://ui.adsabs.harvard.edu/abs/2020arXiv201001224G} {p. arXiv:2010.01224}

\bibitem[\protect\citeauthoryear{{Griffith} \& {Yelle}}{{Griffith} \&
  {Yelle}}{1999}]{GriffithYelle1999}
{Griffith} C.~A.,  {Yelle} R.~V.,  1999, \mn@doi [\apjl] {10.1086/312103},
  \href {https://ui.adsabs.harvard.edu/abs/1999ApJ...519L..85G} {519, L85}

\bibitem[\protect\citeauthoryear{Hansen}{Hansen}{1971}]{Hansen1971}
Hansen J.~E.,  1971, \mn@doi [J. Atmos. Sci.]
  {10.1175/1520-0469(1971)028%3C0120%3AMSOPLI%3E2.0.CO;2}, 28, 120

\bibitem[\protect\citeauthoryear{{Hawley}, {Gizis}  \& {Reid}}{{Hawley}
  et~al.}{1997}]{Hawley1997}
{Hawley} S.~L.,  {Gizis} J.~E.,   {Reid} N.~I.,  1997, \mn@doi [\aj]
  {10.1086/118363}, \href
  {https://ui.adsabs.harvard.edu/abs/1997AJ....113.1458H} {113, 1458}

\bibitem[\protect\citeauthoryear{{Heintz}}{{Heintz}}{1994}]{Heintz1994}
{Heintz} W.~D.,  1994, \mn@doi [\aj] {10.1086/117247}, \href
  {https://ui.adsabs.harvard.edu/abs/1994AJ....108.2338H} {108, 2338}

\bibitem[\protect\citeauthoryear{{Hinkel}, {Timmes}, {Young}, {Pagano}  \&
  {Turnbull}}{{Hinkel} et~al.}{2014}]{hinkel2014}
{Hinkel} N.~R.,  {Timmes} F.~X.,  {Young} P.~A.,  {Pagano} M.~D.,   {Turnbull}
  M.~C.,  2014, \mn@doi [\aj] {10.1088/0004-6256/148/3/54}, \href
  {https://ui.adsabs.harvard.edu/abs/2014AJ....148...54H} {148, 54}

\bibitem[\protect\citeauthoryear{Huffman \& Wild}{Huffman \&
  Wild}{1967}]{HuffmanWild1967}
Huffman D.~R.,  Wild R.~L.,  1967, \mn@doi [Phys. Rev.]
  {10.1103/PhysRev.156.989}, 156, 989

\bibitem[\protect\citeauthoryear{{John}}{{John}}{1988}]{John1988}
{John} T.~L.,  1988, \aap, \href
  {https://ui.adsabs.harvard.edu/abs/1988A&A...193..189J} {193, 189}

\bibitem[\protect\citeauthoryear{Kass \& Raftery}{Kass \&
  Raftery}{1995}]{KassRaferty1995}
Kass R.~E.,  Raftery A.~E.,  1995, \mn@doi [Journal of the American Statistical
  Association] {10.1080/01621459.1995.10476572}, 90, 773

\bibitem[\protect\citeauthoryear{Khachai, Khenata, Bouhemadou, Haddou, Reshak,
  Amrani, Rached  \& Soudini}{Khachai et~al.}{2009}]{Khachai2009}
Khachai H.,  Khenata R.,  Bouhemadou A.,  Haddou A.,  Reshak A.,  Amrani B.,
  Rached D.,   Soudini B.,  2009, \mn@doi [Journal of physics. Condensed matter
  : an Institute of Physics journal] {10.1088/0953-8984/21/9/095404}, 21,
  095404

\bibitem[\protect\citeauthoryear{{Kirkpatrick}}{{Kirkpatrick}}{2005}]{Kirkpatrick2005}
{Kirkpatrick} J.~D.,  2005, \mn@doi [\araa]
  {10.1146/annurev.astro.42.053102.134017}, \href
  {https://ui.adsabs.harvard.edu/abs/2005ARA&A..43..195K} {43, 195}

\bibitem[\protect\citeauthoryear{{Kirkpatrick} et~al.,}{{Kirkpatrick}
  et~al.}{2021}]{Kirkpatrick2021}
{Kirkpatrick} J.~D.,  et~al., 2021, \mn@doi [\apjs] {10.3847/1538-4365/abd107},
  \href {https://ui.adsabs.harvard.edu/abs/2021ApJS..253....7K} {253, 7}

\bibitem[\protect\citeauthoryear{{Lebreton}, {Gomez}, {Mermilliod}  \&
  {Perryman}}{{Lebreton} et~al.}{1997}]{Lebreton1997}
{Lebreton} Y.,  {Gomez} A.~E.,  {Mermilliod} J.~C.,   {Perryman} M.~A.~C.,
  1997, in {Bonnet} R.~M.,  et~al., eds,  ESA Special Publication Vol. 402,
  Hipparcos - Venice '97. pp 231--236

\bibitem[\protect\citeauthoryear{{Line}, {Teske}, {Burningham}, {Fortney}  \&
  {Marley}}{{Line} et~al.}{2015}]{Line2015}
{Line} M.~R.,  {Teske} J.,  {Burningham} B.,  {Fortney} J.~J.,   {Marley}
  M.~S.,  2015, \mn@doi [\apj] {10.1088/0004-637X/807/2/183}, \href
  {https://ui.adsabs.harvard.edu/abs/2015ApJ...807..183L} {807, 183}

\bibitem[\protect\citeauthoryear{{Line} et~al.,}{{Line}
  et~al.}{2017}]{Line2017}
{Line} M.~R.,  et~al., 2017, \mn@doi [\apj] {10.3847/1538-4357/aa7ff0}, \href
  {https://ui.adsabs.harvard.edu/abs/2017ApJ...848...83L} {848, 83}

\bibitem[\protect\citeauthoryear{{Liu} et~al.,}{{Liu} et~al.}{2013}]{liu2013}
{Liu} M.~C.,  et~al., 2013, \mn@doi [\apjl] {10.1088/2041-8205/777/2/L20},
  \href {https://ui.adsabs.harvard.edu/abs/2013ApJ...777L..20L} {777, L20}

\bibitem[\protect\citeauthoryear{{Lodders} \& {Fegley}}{{Lodders} \&
  {Fegley}}{2002}]{LoddersFegley2002}
{Lodders} K.,  {Fegley} B.,  2002, \mn@doi [\icarus] {10.1006/icar.2001.6740},
  \href {https://ui.adsabs.harvard.edu/abs/2002Icar..155..393L} {155, 393}

\bibitem[\protect\citeauthoryear{{Lodders} \& {Fegley}}{{Lodders} \&
  {Fegley}}{2006}]{LoddersFegley2006}
{Lodders} K.,  {Fegley} B. J.,  2006, {Chemistry of Low Mass Substellar
  Objects}.
p.~1, \mn@doi{10.1007/3-540-30313-8_1}

\bibitem[\protect\citeauthoryear{{Madhusudhan} \& {Seager}}{{Madhusudhan} \&
  {Seager}}{2009}]{MadhusudhanSeager2009}
{Madhusudhan} N.,  {Seager} S.,  2009, \mn@doi [\apj]
  {10.1088/0004-637X/707/1/24}, \href
  {https://ui.adsabs.harvard.edu/abs/2009ApJ...707...24M} {707, 24}

\bibitem[\protect\citeauthoryear{{Manjavacas} et~al.,}{{Manjavacas}
  et~al.}{2019}]{Manjavacas2019}
{Manjavacas} E.,  et~al., 2019, \mn@doi [\apjl] {10.3847/2041-8213/ab13b9},
  \href {https://ui.adsabs.harvard.edu/abs/2019ApJ...875L..15M} {875, L15}

\bibitem[\protect\citeauthoryear{{Marley} \& {Robinson}}{{Marley} \&
  {Robinson}}{2015}]{MarleyRobinson2015}
{Marley} M.~S.,  {Robinson} T.~D.,  2015, \mn@doi [\araa]
  {10.1146/annurev-astro-082214-122522}, \href
  {https://ui.adsabs.harvard.edu/abs/2015ARA&A..53..279M} {53, 279}

\bibitem[\protect\citeauthoryear{{Marley} et~al.,}{{Marley}
  et~al.}{2021}]{MarleyNEW2021}
{Marley} M.~S.,  et~al., 2021, arXiv e-prints, \href
  {https://ui.adsabs.harvard.edu/abs/2021arXiv210707434M} {p. arXiv:2107.07434}

\bibitem[\protect\citeauthoryear{{Miles-P{\'a}ez}, {Metchev}, {Luhman},
  {Marengo}  \& {Hulsebus}}{{Miles-P{\'a}ez} et~al.}{2017}]{milespaez2017}
{Miles-P{\'a}ez} P.~A.,  {Metchev} S.,  {Luhman} K.~L.,  {Marengo} M.,
  {Hulsebus} A.,  2017, \mn@doi [\aj] {10.3847/1538-3881/aa9711}, \href
  {https://ui.adsabs.harvard.edu/abs/2017AJ....154..262M} {154, 262}

\bibitem[\protect\citeauthoryear{{Miles} et~al.,}{{Miles}
  et~al.}{2020}]{Miles2020}
{Miles} B.~E.,  et~al., 2020, \mn@doi [\aj] {10.3847/1538-3881/ab9114}, \href
  {https://ui.adsabs.harvard.edu/abs/2020AJ....160...63M} {160, 63}

\bibitem[\protect\citeauthoryear{{Morley}, {Fortney}, {Marley}, {Visscher},
  {Saumon}  \& {Leggett}}{{Morley} et~al.}{2012}]{Morley2012}
{Morley} C.~V.,  {Fortney} J.~J.,  {Marley} M.~S.,  {Visscher} C.,  {Saumon}
  D.,   {Leggett} S.~K.,  2012, \mn@doi [\apj] {10.1088/0004-637X/756/2/172},
  \href {https://ui.adsabs.harvard.edu/abs/2012ApJ...756..172M} {756, 172}

\bibitem[\protect\citeauthoryear{{Morley}, {Marley}, {Fortney}, {Lupu},
  {Saumon}, {Greene}  \& {Lodders}}{{Morley} et~al.}{2014}]{Morley2014}
{Morley} C.~V.,  {Marley} M.~S.,  {Fortney} J.~J.,  {Lupu} R.,  {Saumon} D.,
  {Greene} T.,   {Lodders} K.,  2014, \mn@doi [\apj]
  {10.1088/0004-637X/787/1/78}, \href
  {https://ui.adsabs.harvard.edu/abs/2014ApJ...787...78M} {787, 78}

\bibitem[\protect\citeauthoryear{{Moses}, {Madhusudhan}, {Visscher}  \&
  {Freedman}}{{Moses} et~al.}{2013a}]{Moses2013}
{Moses} J.~I.,  {Madhusudhan} N.,  {Visscher} C.,   {Freedman} R.~S.,  2013a,
  \mn@doi [\apj] {10.1088/0004-637X/763/1/25}, \href
  {https://ui.adsabs.harvard.edu/abs/2013ApJ...763...25M} {763, 25}

\bibitem[\protect\citeauthoryear{{Moses} et~al.,}{{Moses}
  et~al.}{2013b}]{Moses2013B}
{Moses} J.~I.,  et~al., 2013b, \mn@doi [\apj] {10.1088/0004-637X/777/1/34},
  \href {https://ui.adsabs.harvard.edu/abs/2013ApJ...777...34M} {777, 34}

\bibitem[\protect\citeauthoryear{{Nakajima} \& {Sorahana}}{{Nakajima} \&
  {Sorahana}}{2016}]{Nakajima2016}
{Nakajima} T.,  {Sorahana} S.,  2016, \mn@doi [\apj]
  {10.3847/0004-637X/830/2/159}, \href
  {https://ui.adsabs.harvard.edu/abs/2016ApJ...830..159N} {830, 159}

\bibitem[\protect\citeauthoryear{{Nakajima}, {Morino}  \&
  {Fukagawa}}{{Nakajima} et~al.}{2010}]{Nakajima2010}
{Nakajima} T.,  {Morino} J.-I.,   {Fukagawa} M.,  2010, \mn@doi [\aj]
  {10.1088/0004-6256/140/3/713}, \href
  {https://ui.adsabs.harvard.edu/abs/2010AJ....140..713N} {140, 713}

\bibitem[\protect\citeauthoryear{{Nissen}}{{Nissen}}{2013}]{Nissen2013}
{Nissen} P.~E.,  2013, \mn@doi [\aap] {10.1051/0004-6361/201321234}, \href
  {https://ui.adsabs.harvard.edu/abs/2013A&A...552A..73N} {552, A73}

\bibitem[\protect\citeauthoryear{{Noll}, {Geballe}  \& {Marley}}{{Noll}
  et~al.}{1997}]{Noll1997}
{Noll} K.~S.,  {Geballe} T.~R.,   {Marley} M.~S.,  1997, \mn@doi [\apjl]
  {10.1086/310954}, \href
  {https://ui.adsabs.harvard.edu/abs/1997ApJ...489L..87N} {489, L87}

\bibitem[\protect\citeauthoryear{{{\"O}berg}, {Murray-Clay}  \&
  {Bergin}}{{{\"O}berg} et~al.}{2011}]{Oberg2011}
{{\"O}berg} K.~I.,  {Murray-Clay} R.,   {Bergin} E.~A.,  2011, \mn@doi [\apjl]
  {10.1088/2041-8205/743/1/L16}, \href
  {https://ui.adsabs.harvard.edu/abs/2011ApJ...743L..16O} {743, L16}

\bibitem[\protect\citeauthoryear{Querry}{Querry}{1987}]{Querry1987}
Querry M.,  1987.

\bibitem[\protect\citeauthoryear{{Richard} et~al.,}{{Richard}
  et~al.}{2012}]{Richard2012}
{Richard} C.,  et~al., 2012, \jqsrt, 113, 1276

\bibitem[\protect\citeauthoryear{{Rosenthal} et~al.,}{{Rosenthal}
  et~al.}{2021}]{rosenthal2021}
{Rosenthal} L.~J.,  et~al., 2021, \mn@doi [\apjs] {10.3847/1538-4365/abe23c},
  \href {https://ui.adsabs.harvard.edu/abs/2021ApJS..255....8R} {255, 8}

\bibitem[\protect\citeauthoryear{{Saumon} \& {Marley}}{{Saumon} \&
  {Marley}}{2008}]{SaumonMarley2008}
{Saumon} D.,  {Marley} M.~S.,  2008, \mn@doi [\apj] {10.1086/592734}, \href
  {https://ui.adsabs.harvard.edu/abs/2008ApJ...689.1327S} {689, 1327}

\bibitem[\protect\citeauthoryear{{Saumon}, {Marley}, {Cushing}, {Leggett},
  {Roellig}, {Lodders}  \& {Freedman}}{{Saumon} et~al.}{2006}]{Saumon2006}
{Saumon} D.,  {Marley} M.~S.,  {Cushing} M.~C.,  {Leggett} S.~K.,  {Roellig}
  T.~L.,  {Lodders} K.,   {Freedman} R.~S.,  2006, \mn@doi [\apj]
  {10.1086/505419}, \href
  {https://ui.adsabs.harvard.edu/abs/2006ApJ...647..552S} {647, 552}

\bibitem[\protect\citeauthoryear{{Saumon}, {Marley}, {Abel}, {Frommhold}  \&
  {Freedman}}{{Saumon} et~al.}{2012}]{Saumon2012}
{Saumon} D.,  {Marley} M.~S.,  {Abel} M.,  {Frommhold} L.,   {Freedman} R.~S.,
  2012, \mn@doi [\apj] {10.1088/0004-637X/750/1/74}, \href
  {https://ui.adsabs.harvard.edu/abs/2012ApJ...750...74S} {750, 74}

\bibitem[\protect\citeauthoryear{{Schneider}, {Windsor}, {Cushing},
  {Kirkpatrick}  \& {Wright}}{{Schneider} et~al.}{2016}]{schneider2016}
{Schneider} A.~C.,  {Windsor} J.,  {Cushing} M.~C.,  {Kirkpatrick} J.~D.,
  {Wright} E.~L.,  2016, \mn@doi [\apjl] {10.3847/2041-8205/822/1/L1}, \href
  {https://ui.adsabs.harvard.edu/abs/2016ApJ...822L...1S} {822, L1}

\bibitem[\protect\citeauthoryear{{Scott} \& {Duley}}{{Scott} \&
  {Duley}}{1996}]{ScottDuley1996}
{Scott} A.,  {Duley} W.~W.,  1996, \mn@doi [\apjs] {10.1086/192321}, \href
  {https://ui.adsabs.harvard.edu/abs/1996ApJS..105..401S} {105, 401}

\bibitem[\protect\citeauthoryear{{Soderblom}, {Laskar}, {Valenti}, {Stauffer}
  \& {Rebull}}{{Soderblom} et~al.}{2009}]{soderblom2009}
{Soderblom} D.~R.,  {Laskar} T.,  {Valenti} J.~A.,  {Stauffer} J.~R.,
  {Rebull} L.~M.,  2009, \mn@doi [\aj] {10.1088/0004-6256/138/5/1292}, \href
  {https://ui.adsabs.harvard.edu/abs/2009AJ....138.1292S} {138, 1292}

\bibitem[\protect\citeauthoryear{{Stamatellos} \& {Whitworth}}{{Stamatellos} \&
  {Whitworth}}{2009}]{StametellosWhitworth2009}
{Stamatellos} D.,  {Whitworth} A.~P.,  2009, in {Stempels} E.,  ed.,  American
  Institute of Physics Conference Series Vol. 1094, 15th Cambridge Workshop on
  Cool Stars, Stellar Systems, and the Sun. pp 557--560 (\mn@eprint {arXiv}
  {0809.5042}), \mn@doi{10.1063/1.3099172}

\bibitem[\protect\citeauthoryear{{Tannock}, {Metchev}, {Hood}, {Mace},
  {Fortney}, {Morley}, {Jaffe}  \& {Lupu}}{{Tannock}
  et~al.}{2022}]{Tannock2022}
{Tannock} M.~E.,  {Metchev} S.,  {Hood} C.~E.,  {Mace} G.~N.,  {Fortney} J.~J.,
   {Morley} C.~V.,  {Jaffe} D.~T.,   {Lupu} R.,  2022, \mn@doi [\mnras]
  {10.1093/mnras/stac1412}, \href
  {https://ui.adsabs.harvard.edu/abs/2022MNRAS.514.3160T} {514, 3160}

\bibitem[\protect\citeauthoryear{{Theissen}, {Burgasser}, {Bardalez Gagliuffi},
  {Hardegree-Ullman}, {Gagn{\'e}}, {Schmidt}  \& {West}}{{Theissen}
  et~al.}{2018}]{theissen2018}
{Theissen} C.~A.,  {Burgasser} A.~J.,  {Bardalez Gagliuffi} D.~C.,
  {Hardegree-Ullman} K.~K.,  {Gagn{\'e}} J.,  {Schmidt} S.~J.,   {West} A.~A.,
  2018, \mn@doi [\apj] {10.3847/1538-4357/aaa0cf}, \href
  {https://ui.adsabs.harvard.edu/abs/2018ApJ...853...75T} {853, 75}

\bibitem[\protect\citeauthoryear{{Tsuji} \& {Nakajima}}{{Tsuji} \&
  {Nakajima}}{2014}]{TsujiNakajima2014}
{Tsuji} T.,  {Nakajima} T.,  2014, \mn@doi [\pasj] {10.1093/pasj/psu078}, \href
  {https://ui.adsabs.harvard.edu/abs/2014PASJ...66...98T} {66, 98}

\bibitem[\protect\citeauthoryear{{Tsuji} \& {Nakajima}}{{Tsuji} \&
  {Nakajima}}{2016}]{TsujiNakajima2016}
{Tsuji} T.,  {Nakajima} T.,  2016, \mn@doi [\pasj] {10.1093/pasj/psv119}, \href
  {https://ui.adsabs.harvard.edu/abs/2016PASJ...68...13T} {68, 13}

\bibitem[\protect\citeauthoryear{{Visscher}}{{Visscher}}{2012}]{Visscher2012}
{Visscher} C.,  2012, \mn@doi [\apj] {10.1088/0004-637X/757/1/5}, \href
  {https://ui.adsabs.harvard.edu/abs/2012ApJ...757....5V} {757, 5}

\bibitem[\protect\citeauthoryear{{Visscher} \& {Moses}}{{Visscher} \&
  {Moses}}{2011}]{VisscherMoses2011}
{Visscher} C.,  {Moses} J.~I.,  2011, \mn@doi [\apj]
  {10.1088/0004-637X/738/1/72}, \href
  {https://ui.adsabs.harvard.edu/abs/2011ApJ...738...72V} {738, 72}

\bibitem[\protect\citeauthoryear{{Visscher}, {Lodders}  \& {Fegley}}{{Visscher}
  et~al.}{2006}]{Visscher2006}
{Visscher} C.,  {Lodders} K.,   {Fegley} Bruce J.,  2006, \mn@doi [\apj]
  {10.1086/506245}, \href
  {https://ui.adsabs.harvard.edu/abs/2006ApJ...648.1181V} {648, 1181}

\bibitem[\protect\citeauthoryear{{Visscher}, {Lodders}  \& {Fegley}}{{Visscher}
  et~al.}{2010}]{Visscher2010}
{Visscher} C.,  {Lodders} K.,   {Fegley} Bruce J.,  2010, \mn@doi [\apj]
  {10.1088/0004-637X/716/2/1060}, \href
  {https://ui.adsabs.harvard.edu/abs/2010ApJ...716.1060V} {716, 1060}

\bibitem[\protect\citeauthoryear{{West}, {Hawley}, {Bochanski}, {Covey},
  {Reid}, {Dhital}, {Hilton}  \& {Masuda}}{{West} et~al.}{2008}]{West2008}
{West} A.~A.,  {Hawley} S.~L.,  {Bochanski} J.~J.,  {Covey} K.~R.,  {Reid}
  I.~N.,  {Dhital} S.,  {Hilton} E.~J.,   {Masuda} M.,  2008, \mn@doi [\aj]
  {10.1088/0004-6256/135/3/785}, \href
  {https://ui.adsabs.harvard.edu/abs/2008AJ....135..785W} {135, 785}

\bibitem[\protect\citeauthoryear{Whitworth}{Whitworth}{2018}]{Whitworth2018}
Whitworth A.~P.,  2018, Brown Dwarf Formation: Theory.
Springer International Publishing, Cham, pp 1--22,
  \mn@doi{10.1007/978-3-319-30648-3_95-1}, \url
  {https://doi.org/10.1007/978-3-319-30648-3_95-1}

\bibitem[\protect\citeauthoryear{{Whitworth}, {Bate}, {Nordlund}, {Reipurth}
  \& {Zinnecker}}{{Whitworth} et~al.}{2007}]{whitworth2007}
{Whitworth} A.,  {Bate} M.~R.,  {Nordlund} {\r{A}}.,  {Reipurth} B.,
  {Zinnecker} H.,  2007, in {Reipurth} B.,  {Jewitt} D.,   {Keil} K.,  eds,
  Protostars and Planets V. p.~459

\bibitem[\protect\citeauthoryear{{Zalesky}, {Line}, {Schneider}  \&
  {Patience}}{{Zalesky} et~al.}{2019}]{Zalesky2019}
{Zalesky} J.~A.,  {Line} M.~R.,  {Schneider} A.~C.,   {Patience} J.,  2019,
  \mn@doi [\apj] {10.3847/1538-4357/ab16db}, \href
  {https://ui.adsabs.harvard.edu/abs/2019ApJ...877...24Z} {877, 24}

\bibitem[\protect\citeauthoryear{{Zalesky}, {Saboi}, {Line}, {Zhang},
  {Schneider}, {Liu}, {Best}  \& {Marley}}{{Zalesky}
  et~al.}{2022}]{Zalesky2022}
{Zalesky} J.~A.,  {Saboi} K.,  {Line} M.~R.,  {Zhang} Z.,  {Schneider} A.~C.,
  {Liu} M.~C.,  {Best} W. M.~J.,   {Marley} M.~S.,  2022, \mn@doi [\apj]
  {10.3847/1538-4357/ac786c}, \href
  {https://ui.adsabs.harvard.edu/abs/2022ApJ...936...44Z} {936, 44}

\bibitem[\protect\citeauthoryear{{Zhang}, {Liu}, {Marley}, {Line}  \&
  {Best}}{{Zhang} et~al.}{2020}]{Zhang2020}
{Zhang} Z.,  {Liu} M.~C.,  {Marley} M.~S.,  {Line} M.~R.,   {Best} W. M.~J.,
  2020, arXiv e-prints, \href
  {https://ui.adsabs.harvard.edu/abs/2020arXiv201112294Z} {p. arXiv:2011.12294}

\makeatother
\end{thebibliography}

\end{document}